\documentclass[11pt]{article}

\newfont{\tenmsb}{msbm10 scaled\magstep1}

\let\ssection=\section\renewcommand{\section}{\setcounter{equation}{0}\ssection}

\usepackage{graphics,amssymb,amsmath,amsthm,amscd,array}
\usepackage{graphicx,subfigure}
\usepackage{makeidx}
\usepackage{bm}% bold math
\usepackage{amsmath,dsfont}
\usepackage{rotating}

\font\BBBig=cmr10 scaled\magstep4
\font\small=cmr9

\def\parag{\hfil\break} %%%%% paragraph
\def\kikezd{\parag\underbar}
\def\IR{{\mathds{R}}} %%%%% Reals
\def\IC{{\mathds{C}}} %%%%% Complex
\def\IZ{{\mathds{Z}}} %%%%% Integers
\def\IS{{\mathds{S}}}
\def\II{{\mathds{1}}}
\def\IT{{\mathds{T}}}
\def\smallover#1/#2{\hbox{$\textstyle\frac{#1}{#2}$}} %
\def\smallcirc{{\,\raise 0.5pt \hbox{$\scriptstyle\circ$}\,}}
\def\2{{\smallover1/2}}
\def\={{\!=\!}}
\def\Ort{{\rm O}}
\def\ort{{\rm o}}
\def\UN{{\rm U}}
\def\un{{\rm u}}
\def\SO{{\rm SO}}
\def\SU{{\rm SU}}
\def\su{{\rm su}}
\def\so{{\rm so}}
\def\SP{{\rm Sp}}

\def\p{{\partial }}
\def\dAlembert{\vcenter {
    \hbox {\vrule height8pt width0.4pt depth0.0pt
           \vrule height8pt width7.2pt depth-7.6pt
           \vrule height8pt width0.4pt depth0.0pt
           \kern-8pt
           \vrule height0.4pt width8pt depth0.0pt
          \,}}}%--- Francisco's box

\def\bp{\bar{\partial}}
\def\bz{{\overline{z}}}
\def\and{\qquad\hbox{\small and}\qquad}

\def\bJ{{\bm{J}}}
\def\bA{{\bm{A}}}
\def\bB{{\bm{B}}}
\def\bD{{\bm{D}}}
\def\bd{{\bm{d}}}
\def\bb{{\bm{b}}}
\def\ba{{\bm{a}}}
\def\bx{{\bm{x}}}
\def\bL{{\bm{L}}}
\def\bM{{\bm{M}}}
\def\bS{{\bm{S}}}

\def\btau{\mbox{\boldmath$\tau$}}
\def\bnabla{\mbox{\boldmath$\nabla$}}
\def\bpar{\mbox{\boldmath$\partial$}}

\def\Dir{
{D\mkern-2mu\llap{{\raise+0.5pt\hbox{\big/}}}\mkern+2mu}
}  %Diracop

\def\oQ{{\stackrel{\smallcirc}{Q}}}
\def\oW{{\stackrel{\smallcirc}{W}}}
\def\oalpha{{\stackrel{\smallcirc}{\alpha}}}
\def\beq{\begin{equation}}
\def\eeq{\end{equation}}
\def\beqa{\begin{eqnarray}}
\def\eeqa{\end{eqnarray}}
\def\nn{\nonumber}
\def\barr{\left(\begin{array}}
\def\earr{\end{array}\right)}
\newcommand{\gk}{\mathfrak{k}}
\newcommand{\gl}{\mathfrak{l}}
\newcommand{\gh}{\mathfrak{h}}
\newcommand{\gt}{\mathfrak{t}}
\def\tr{{\,\rm Tr\,}}
\def\wK{{\widetilde{K}}}
\def\bbeta{{\bm{\beta}}}

\begin{document}
\setlength{\baselineskip}{15pt}

%%%%%%%%%%%%%%%%%%%%%%%%%%%%%%%%%%%%%%%%%%%
%%%%%%%%%%%%%%%% the text %%%%%%%%%%%%%%%%%
%%%%%%%%%%%%%%%%%%%%%%%%%%%%%%%%%%%%%%%%%%%
\title{\BBBig Monopole-charge instability\footnote{Published as \textit{International Journal of Modern Physics} {\bf A3} (1988) 665-702.
}
\\[18pt]}

\author{
Peter A. HORVATHY
\footnote{Present address: Laboratoire de Math\'ematiques et de Physique Th\'eorique,
Universit\'e de TOURS (France).
e-mail: horvathy-at-univ-tours.fr}
\\[10pt]
Centre de Physique Th\'eorique, CNRS Luminy
\\[10pt]  MARSEILLE (FRANCE)
\\[20pt]
L. O'RAIFEARTAIGH\footnote{deceased.}
\\[10pt]
Dublin Institute for Advanced Study
\\[10pt]
DUBLIN (Ireland)
\\[20pt]
J. RAWNSLEY\footnote{e-mail: J.Rawnsley@warwick.ac.uk}
\\[10pt]
The Mathematics Institute, Warwick University
\\[10pt] 
COVENTRY CV4 7AL,
 (England)
}

\maketitle

\begin{abstract}
For monopoles with non-vanishing Higgs potential it is shown that with respect to ``Brandt-Neri-Coleman type'' variations (a) the stability problem reduces to that of a pure gauge theory on the two-sphere
(b) 
each topological sector admits one, and only one, stable monopole charge, and
(c) each unstable monopole admits $\displaystyle{2\sum_{q<0} \left(2|q|-1\right)}$
negative modes, where the sum goes over all negative eigenvalues $q$ of the non-Abelian charge $Q$. An explicit construction for 
(i) the unique stable charge (ii) the negative modes and (iii) the spectrum of the Hessian,
on the $2$-sphere, is then given. The relation to loops in the residual group is explained. The negative modes are tangent to suitable energy-reducing two-spheres. The general theory is illustrated for the little groups $\UN(2), \UN(3),
\SU(3)/\IZ_3$ and $\Ort(5)$.
\end{abstract}
\vskip15mm
e-print: \texttt{arXiv:0909.2523 [hep-th]}

\newpage
%\null\newpage
\tableofcontents
%\newpage\null\newpage

%%%%%%%%%%%%%%%%%%%%%%
\section{INTRODUCTION}
%%%%%%%%%%%%%%%%%

By linearizing the field equations around a monopole solution, Brandt and Neri \cite{1} and Coleman \cite{2}
 have shown that most non-Abelian monopoles are \textit{unstable}
 with respect to small perturbations, unless all eigenvalues, $q$, of the non-Abelian charge,
 $Q$, (Ref. \cite{3}) satisfy the ``Brandt-Neri condition''
\beq
q=0\quad
\hbox{or}\quad
\pm\2\, .
\label{1.1}
\eeq
Goddard and Olive \cite{4} prove then that the semisimple part of 
$Q$ must be of a very special form, known in representation theory as a ``minimal vector'' or a
``minimal co-weight"" (see Secs. 2 and 4 for details).

Asymptotic monopoles with residual group $H$ behave very much like pure
Yang-Mills theory on $\IS^2$ with gauge group $H$. The solutions of the  Yang-Mills (YM) equations are again characterized by a $Q$. But YM on $\IS^2$ is just a special case of YM on a Riemann surface, studied by Atiyah and Bott \cite{6}. It follows then from the general theory that most solutions are unstable, and admit rather 
\beq
\nu=2\sum_{q\,<\,0} \left(2|q|-1\right)
\label{1.2}
\eeq
negative modes, where the sum goes over all negative eigenvalues $q<0$ of $Q$ \cite{5,7}.
Note that $2q$ is always an integer because $2Q$ is a charge (see below). The zero eigenvalues do not appear in the sum in (\ref{1.2}) and the eigenvalues $q=\pm\2$ do not contribute.
(\ref{1.2}) is hence consistent with the Brandt-Neri condition (\ref{1.1}).

The aim of this paper is to relate and complete the above results. After summarizing
the necessary algebraic tools (and in particular the basic properties of minimal co-weights), we review those properties of finite-energy configurations (Sec. 3) and of solutions (Sec. 4) which are relevant for our purposes. Much of the content of these
sections is already known \cite{2,4,8} but we have assembled the results from different sources and summarized them for completeness and for the convenience of the reader.

As well-known, monopoles fall into topological sectors separated by infinite energy barriers and labelled by homotopy classes in $\pi_1$ of $H$, the residual group after spontaneous symmetry breaking. We show that, for any compact and connected $H$,
each topological sector contains a unique ``minimal'' charge, $\oQ$, i.e. one whose semi-simple part is a minimal co-weight. Minimal charges are thus introduced here independently of stability considerations, as \textit{labels of the topological sectors}.

Monopoles are critical points of the Yang-Mills energy functional. Our first
approach to instability is \textit{local} in the sense that it only involves the behaviour of the energy functional in the neighbourhood of a critical point. This behaviour is characterized by the second variation, called the Hessian. More generally, if a field $\psi$ is a critical
point of some energy functional $E(\psi)$, then $\delta E(\delta\psi)=0$ for any
variation $\delta\psi$. The expansion
\beq
E(\psi+\delta\psi)=E(\psi)+\2\delta^2E\big(\delta\psi,\delta\psi\big)+{\Ort}\Big(\delta\psi^3\Big)
\label{1.3}
\eeq
shows then that $\psi$ is locally stable if the Hessian has no negative eigenvalues. Having
 a negative value would mean in fact that the excitation $\delta\psi$  has negative mass-square, i.e. the energy of the configuration $\psi$ could be reduced by tachyon formation.
 
\begin{figure}
\begin{center}\hspace{-6mm}
\includegraphics[scale=0.45]{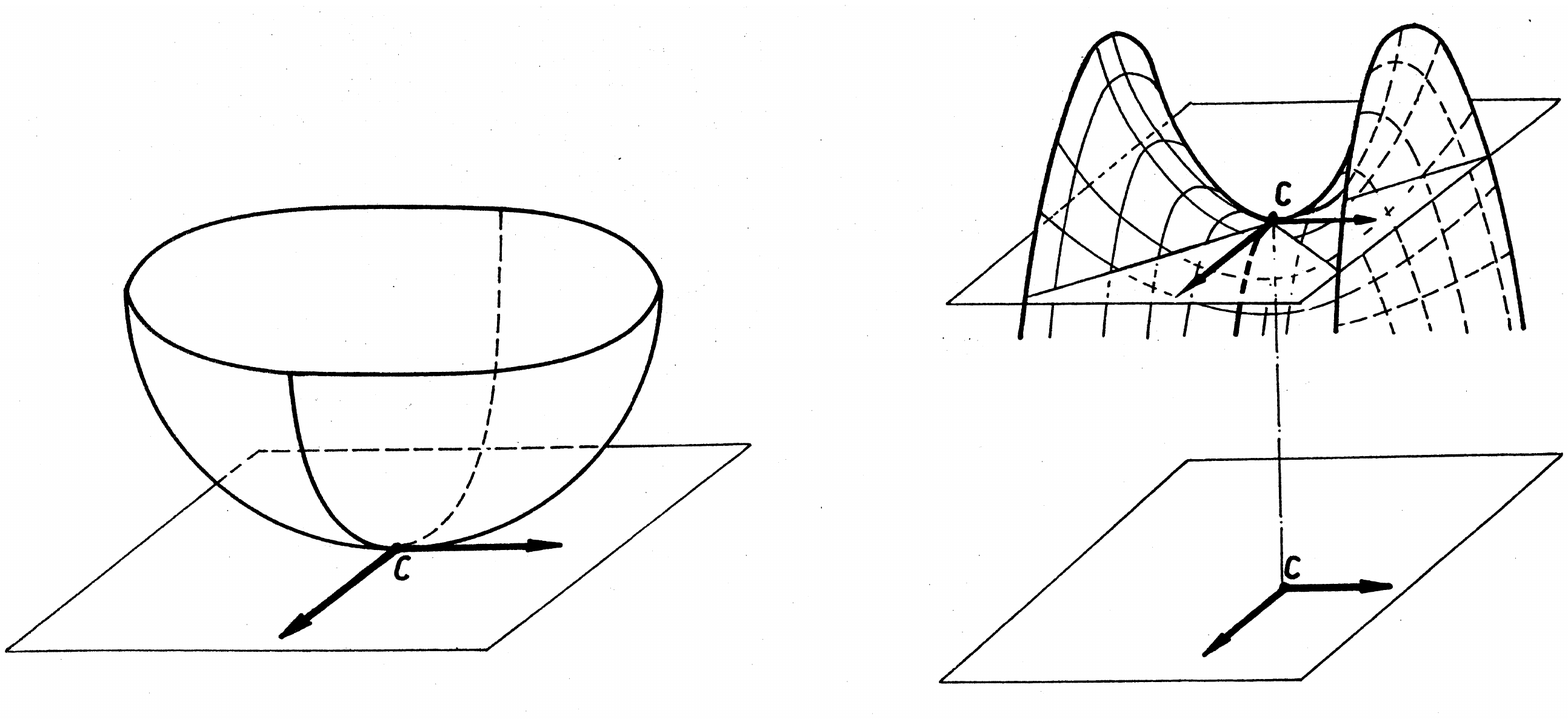}\vspace{-26mm}
\end{center}
\caption{\it The energy functional is a surface over finite energy field configurations. Monopoles are critical points whose local stability depends on the shape of the surface in the neighbourhood of the critical point. For example, the critical point on Fig. 1a is stable, while that on Fig. 1b is unstable.}
\label{Fig1}
\end{figure}

 Geometrically, the energy is an (infinite-dimensional) surface over finite-energy
 configurations, and the stability of a critical point depends on the shape of the
 surface, cf. Fig. \ref{Fig1}.
 
In this paper we restrict our attention to those asymptotic variations of the gauge field alone previously considered by Brandt and Neri \cite{1} and by Coleman \cite{2}. We show that for non-zero Higgs potentials the $3$-dimensional problem essentially reduces to pure YM theory on $\IS^2$.

Other types of variations may also lead to instability. For example, a multicharged configuration can dissociate into single monopoles \cite{9}. The non-Abelian
charge $Q$ is kept fixed under such a process. Our problem here is therefore different:
we inquire about the stability of the charge $Q$ itself. More precisely, we want to know whether a configuration with charge 
$Q_1$ can decay into another one, whose charge is
$Q_2$. Thus the relevance of the problem studied in Ref.~\cite{9} is that, in multiply-charged topological sectors, the vacuum itself may not exist i.e. there are no static solutions to the
field equations (the energy has an infimum but no minimum). Such a situation would, of course, reduce the importance of our results. It seems, though, to be
rather exceptional \cite{10}.

Another interesting type of monopole instability is the one studied by Taubes \cite{11},
whose results concern Prasad-Sommerfield monopoles. They correspond to
variations of the Higgs field rather than to those of the gauge field. Our results here are
complementary to these aspects.

For $\IS^2$, the general theory of Atiyah and Bott
\cite{6} can be related to the
Brandt-Neri-Coleman rotation-group approach. Indeed, on the $q$-eigenspace the interesting part
of the Hessian is
\beq
\int drd\Omega\,\tr\Big\{\big\{(\bJ^2-q(q+1))
\delta\bA\big\}\delta\bA\Big\}+q\!\int drd\Omega\, \tr
(\delta\bA)^2,
\label{1.4}
\eeq
where $\bJ^2=j(j+1)$ is the Casimir of the angular momentum vector
$\bJ$ of the spin-$1$ field $\delta\bA$. Since the first term
is non-negative and the first non-zero eigenvalue is at least $2|q|$, a negative mode can
occur only if the first term in
(\ref{1.4}) vanishes and the second is negative, which only happens if
\beq
q\leq-1
\qquad\hbox{and}\qquad
j=|q|-1.
\label{1.5}
\eeq
From this result it is evident that the negative modes form a $2j+1=2|q|-1$
dimensional $\SU(2)$ multiplet. A simple way of counting the number of
negative modes is to use the diagram introduced by Bott \cite{12}.

The special form (\ref{1.4}) of the Hessian makes it possible
to construct the negative modes explicitly: in terms of the complex (stereographic) coordinates $z$ and
$\bz$ on $\IS^2$, they are given
\beqa
\left(\begin{array}{c}
a_{\bz}
\\
0
\end{array}\right)
=
\barr{c}
\bz^k(1+z\bz)^{-|q|}E_\alpha
\\
0
\earr,
\qquad
\barr{c}0\\
a_z
\earr=
\barr{c}
0\\
z^k(1+z\bz)^{-|q|}E_{-\alpha}
\earr
\label{1.6}
\eeqa
$k=0,\dots,2|q|-2$, where the $E_\alpha$'s are those eigenvectors of $\big[Q,\,\cdot\,\big]$
with eigenvalues $q=\alpha(Q)\leq-1$.
The positive modes ($j\geq|q|$ states) may be constructed by the same technique.

In the Brandt-Neri case $q=0$ or $\pm\2$ there are 
\emph{no}
$j=|q|-1$ states, and the monopole is \emph{stable}.
It follows from the topological  formulation that, 
for any compact $H$, the only charge which satisfies this condition is $Q$ itself. Physically, in each topological sector, $Q$ minimizes the energy in the Coulomb tail \cite{4}.

The integrand in the Hessian is essentially a 
\emph{supersymmetric  Hamiltonian} on $\IS^2$, and the negative modes correspond to its ground state, whose multiplicity (called the
\textit{Witten index}) is exactly the instability index $2|q|-1$ \cite{13}. This is also the
\textit{Atiyah-Singer index} for vectors on $\IS^2$.

For Bogomolny-Prasad-Sommerfield monopoles \cite{14} there is an extra
term $q^2$ in the Hessian due to the long-range Higgs field, which cancels the corresponding term in
(\ref{1.4}) and the relevant part of the Hessian is rather
\beq
\int dr d\Omega\tr (\bJ^2\delta\bA\delta\bA),
\label{1.7}
\eeq
which is manifestly positive. It follows that BPS monopoles are \emph{stable} with respect to
variations of the gauge field alone. (See, however, Ref. \cite{11}).

Another intuitive way of understanding monopole instability is by thinking of them as \textit{elastic strings} \cite{2}: 
\emph{monopoles decay just like strings shrink to shorter configurations} (actually to the shortest one allowed by the topology). 
Remarkably, this analogy can be made rigorous. Indeed, the well-known expression
\beq
h^A(\varphi)={\cal P}\left(\exp\oint_{\gamma_{\varphi}}\bA\right)\,,
\label{1.8}
\eeq
where $\gamma_{\varphi}(\theta)$ is a $1$-parameter family of loops sweeping through the two-sphere, associates
a loop in the residual group $H$ to any YM potential $\bA$ on $\IS^2$.

The map (\ref{1.8}) has been used before \cite{2,3,8} for describing the topological sector
of a monopole. It contains however much more information: as a matter of fact,
it puts \emph{all} homotopy groups of finite-energy YM configurations and of loops in
$H$ in a (1-1) correspondence \cite{15}.

The energy of a loop in $H$ can be defined (Sec. 7) and a variational calculus,
analogous to YM on $\IS^2$, can be developed (this is in fact a kind of ``$1$-dimensional
$\sigma$-model'). Remarkably, the map (\ref{1.8}) carries monopoles i.e. critical points of the YMH functional, into geodesics, which are critical points of the loop-energy functional. Furthermore,
the number of instabilities is also the same, namely (\ref{1.2}) \cite{7}.

These facts are explained by Morse theory \cite{16}: the energy functionals of both YM on $\IS^2$ and of loops in $H$ are ``perfect Morse functions'', and so their
critical points correspond to changes in the topology of the underlying space \cite{16}. But the map (\ref{1.8}) is
a homotopy equivalence \cite{15}, so all topological properties of the two spaces are the same.

A convenient choice of the $\gamma_{\varphi}(\theta)$'s allows also to recover the loop-negative modes (explicitly constructed in Sec. 7) as images of the YM-modes (\ref{1.6}).

After these local considerations
 we investigate the \emph{global} properties.
 What happens in fact to an unstable monopole~? Although it cannot leave its topological sector, it can
 go into another state in the same sector, because all such configurations are separated only by finite energy.

\begin{figure}
\begin{center}\hspace{-2mm}
\includegraphics[scale=0.5]{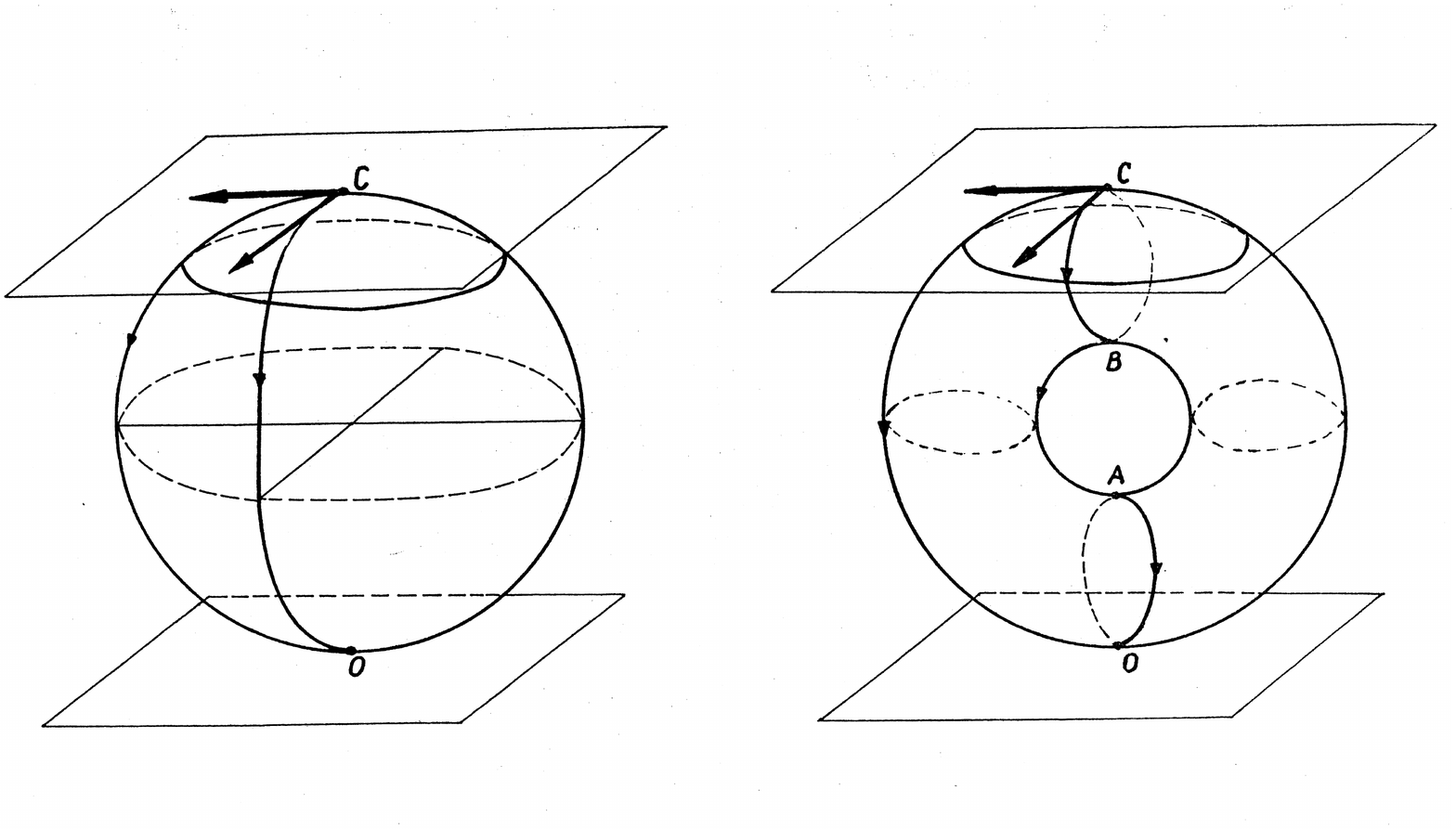}\vspace{-22mm}
\end{center}
\caption{\it Global aspects of instability. A ball put to the top of a sphere (Fig.2a) or to that of a torus
(Fig.2b) rolls down to another, lower-lying critical point and ultimately arrives to the stable configuration.
The $\nu=2$ critical points correspond to non-vanishing classes in $H_2$, the second homology group.
$H_2=\pi_2=\IZ$ for $\IS^2$, while $H_2=\IZ$ and $\pi_2=0$ for the torus.}
\label{Fig2}
\end{figure}

Semiclassically, an unstable monopole will move as to decrease 
its energy.
For example, a ball put to the top of a torus will roll down to another critical point (Fig. \ref{Fig2}). If this is again an unstable configuration, it will continue to roll until
 it arrives at a stable position. 

%%%%%%%%%%
%%%%%%

Most saddle-points in field theories studied so far are associated to \emph{noncontractible} (hyper)\emph{loops} of field configurations \cite{17}. It is easy to see that there are no non-contractible loops in our case. There are, however, non-contractible \emph{spheres}. In Sec.
8 we construct \emph{energy reducing two-spheres} between a given unstable configuration and
certain other, lower-energy configurations. The number of independent two-spheres  is half the number $\nu$ in (\ref{1.2}), and their tangent vectors at the top yield negative modes (cf. Fig. \ref{Fig2}). We also hope that our energy-reducing spheres provide some information on the
possible \textit{routes of decay} for the monopole.

Section 9 is devoted to examples. First we study the residual group $H=\UN(2)$
\cite{18} and $H=U(3)$. Another nice example is provided by $H=\SU(3)/\IZ_3$, previously
studied in higher-dimensional Yang-Mills theory \cite{19}. A simple example where the
special property of the stable charges (mentioned above) enters, is when the
semisimple part $H$ is (a covering of) $\SO(5)$.

%%%%%%%%%%%%%%%%%%%%%%%%%%%%%%%%%
\section{ALGEBRAIC STRUCTURE}
%%%%%%%%%%%%%%%%%%%%%%%%%%%%%%%%%%%

Let us consider a compact simple Lie algebra $\gk$ and
choose a Cartan subalgebra $\gl$. A root $\alpha$ is a linear function on the complexified Cartan algebra 
$\gl^{\IC}$, and to each $\alpha$ is associated a vector $E_\alpha$ (the familiar step operator) from
$\gk^{\IC}$ which satisfies, with any vector $H$ from
$\gl^{\IC}$, the relation
\beq
\big[H,E_\alpha\big]=\alpha(H)E_\alpha\, .
\label{2.1}
\eeq
There exists a set of primitive roots $\alpha_i,\, i=
1,\dots,r$ ($r =$ rank) such that every positive root is a linear combination of the $\alpha_i$ with non-negative integer coefficients
i.e. $\alpha=\sum m_i\alpha_i$ for all $\alpha$.

Alternatively, we can consider the real combinations 
$X_\alpha=E_\alpha+E_{-\alpha}$ and $Y_\alpha=-i(E_\alpha-
E_{-\alpha})$ which satisfy $[H,X_\alpha]=iq_\alpha Y_\alpha$,
$[H,Y_\alpha]=-iq_\alpha X_\alpha$, ($q_\alpha=\alpha(H)$).

If $\alpha$ is a root, define the vector $H_\alpha$ in $\gl^{\IC}$
by $\alpha(X)=\tr(H_\alpha X)$. Choosing the normalization 
$\tr(E_\alpha,E_{-\alpha})=1$, we have 
$[E_\alpha,E_{-\alpha}]=H_\alpha$, 
$[X_\alpha,Y_\alpha]=2iH_\alpha$. Therefore, for each
root $\alpha,\, H_\alpha$ and the $E_{\pm\alpha}$'s (or the real
combinations $X_\alpha$ and $Y_\alpha$) form
 $\SO(3)$ subalgebras of $\gk$.

The \textit{primitive charges} $Q_i$ are defined by
\beq
Q_i=\frac{2H_i}{\tr(H_i^2)}
\quad
\hbox{where}\quad H_i=H_{\alpha_i}.
\label{2.2}
\eeq

The primitive charges form a natural (nonorthogonal) basis for the
Cartan algebra and by adding the $E_\alpha$'s we get a basis for the Lie algebra $\gk^{\IC}$. Similarly, the primitive charges and the
$\big\{\,X_\alpha, Y_\alpha\,\big\}$ form a basis for the real algebra $\gk$.
The integer combinations
$\sum_in_iQ_i$ of the primitive charges form an $r$-dimensional lattice $\Gamma_P$ sitting in the Cartan algebra.

Let us introduce next another basis for the Cartan algebra with elements
$W_i$ dual to the primitive roots,
\beq
\alpha_i(W_j)=\tr(H_iW_j)=\delta_{ij},
\qquad
i,j=1,\dots,r.
\label{2.3}
\eeq
Comparing (\ref{2.3}) with the conventional definition \cite{15}
of primitive weights, for which there is an extra factor $(\alpha_i,\alpha_i)/2$ in front of the $\delta_{ij}$, one sees that the $W_i$'s are just re-scaled weights. They are called 
co-weights \cite{4} and it is evident that they can be normalized so as to coincide with the conventional weights (by choosing 
$(\alpha_i,\alpha_i)=2$) for all groups whose roots are of the same length, i.e. all groups except
$\SP(2n), \SP(2n+1)$ and $G_2$.

The integer combinations $\sum m_iW_i$ form another lattice we denote by $\Gamma_W$. Since
$\alpha(Q_i)$ is always an integer, the $W$-lattice actually contains the primitive-charge lattice, $\Gamma_P\subset\Gamma_W$. The \emph{root planes} of $\gk$ are those vectors $X$ in the Cartan algebra for which $\alpha(X)$ is an integer, i.e., those vectors which have integer eigenvalues in the adjoint
representation. The root planes intersect in the points of the
$W$-lattice.

Both lattices $\Gamma_P$ and $\Gamma_W$ depend on the Lie algebra and not on the group
it generates. Now we define a third lattice, which does
depend on the global structure.

Denote by $\wK$ the (unique) compact, simple, and simply connected Lie group generated by $\gk$. Any other group $K$ whose Lie algebra is
$\gk$ is then of the form $K=\wK/C$, where $C$ is a subgroup
of $Z=Z(\wK)$, the center of $\wK$. $Z$ is finite and Abelian, so $C$ is always discrete. Since $\wK$ is simply connected, $C$ is just $\pi_1(K)$, the first homotopy group of $K$.

The primitive charges satisfy the quantization condition
 $\widetilde{\exp}2\pi iQ_i=1$  (exponential in $\wK$) and thus also in any
 representation of $\wK$ i.e. in any other group $K$
 with the same Lie algebra. For any set $n_i$, $i=1,\dots,r$ of integers,
\beq
\exp\left[2\pi it\sum n_iQ_i\right],
\qquad
0\leq t\leq 1,
\label{2.4}
\eeq
(exponential in $K$) is hence a contractible loop in all representations. Since any loop is homotopic to one of the form $\exp2\pi i tQ,\ 0\leq t\leq1$, we conclude that the lattice
$\Gamma_P$ consists of the generators of \emph{contractible loops}.

More generally, let us fix a group $K$ (i.e., a representation of $\wK$) and define a general charge $Q$ to be an element of the Cartan algebra such that
\beq
\exp[2\pi iQ]=1
\quad\hbox{in}\quad
K,
\label{2.5}
\eeq
so that $\exp[2\pi itQ],\ 0\leq t\leq 1$, is a loop.

Those $Q$'s satisfying the quantization condition (\ref{2.5}) form the \emph{charge lattice}, denoted by $\Gamma_Q$. It depends on the global structure, but it always contains
$\Gamma_P$, the lattice of contractible loops. $\Gamma_P$ and $\Gamma_Q$ are actually the same for the covering group $\wK$.
More generally, two loops $\exp[2\pi itQ_1]$ and
 $\exp[2\pi itQ_2]$ are homotopic if and only if $Q_1-Q_2$ belongs to $\Gamma_P$, so that $\pi_1(K)$ is the quotient of the lattices $\Gamma_Q$ and $\Gamma_P$.

On the other hand, the charge lattice $\Gamma_Q$ is contained 
in the $W$-lattice $\Gamma_W$, because for any root $\alpha$ and charge $Q$,
\beqa
1&=&\big(\exp[2\pi iE_\alpha]\big)
\big(\exp[2\pi iQ]\big)\big(\exp[-2\pi iE_\alpha]\big)
=
\exp\left[2\pi i\big(e^{2\pi iE_\alpha}Q e^{-2\pi iE_\alpha}\big)
\right]\nn
\\[8pt]
&=&e^{2\pi i\alpha(Q)}\exp[2\pi iE_{\alpha}]
=e^{2\pi i\alpha(Q)},
\nn
\eeqa
and hence $\alpha(Q)$ is an integer.

The three lattices introduced above satisfy therefore the relation
\beq
\Gamma_P\subset\Gamma_Q\subset\Gamma_W.
\label{2.6}
\eeq
In general, $\widetilde{\exp}[2\pi iW_j]$ is not unity in the fundamental representation of $\wK$. It is however unity in the
adjoint representation,
\beq
\widetilde{\exp}[2\pi iW_j]=z_j
\label{2.7}
\eeq
belongs therefore to the \emph{center} of $\wK$. Hence the two lattices
$\Gamma_P$ and $\Gamma_W$ coincide for the adjoint group.

Note that the correspondence $W_j\sim z_j$ is one-to-one only
for $\SU(n)$ since for the other groups there are $r$ $W$'s
but less than $r$ elements in the center (as shown in Table 1).

\begin{sidewaystable}
%\centering\caption{\it The simply connected simple compact Lie groups with non-trivial centers, their minimal co-weights, expressed as matrices and as primitive weights, the representations characterized by co-weights and the expansions of the highest roots in terms of the primitive roots. Here $\sigma_2$ and $\sigma_3$ denote Pauli matrices, $\gamma_\mu$ Clifford matrices, ${\rm y}$ the $\SU(3)$ hypercharge diag$(2,-1_,-1)$ and $\gamma=\gamma_1\dots\gamma_{4n}$.
%\label{rotfloat1}}
\begin{tabular}{|l|l|l|l|l|l|}
%&&&&&\\
\hline
center
&group
&$\oW$ as matrix
&$\oW$ as weight
&$\oW$ representation
&highest root expansion
\\
\hline\hline
$\IZ_2$
&Spin$(2n+1)$
&$\sigma_{\mu\nu}=(i/4)[\gamma_\mu,\gamma_\nu],\,\mu\neq\nu$ 
&$\omega_1$
&$F_1$ (\hbox{vector})
&$\alpha_1+2\alpha_2+\dots+2\alpha_{r-1}+2\alpha_r$
\\
\hline
$\IZ_2$
&Sympl$(2n)$
&$(1/2)\sigma_3\times\II_n$
&$2\omega_r$
&$F_r$ (rank $r$ antisym. tensor)
&$2\alpha_1+2\alpha_2+\dots+2\alpha_{r-1}+\alpha_r$
\\
\hline
$\IZ_2$
&$E_7$
&$(1/2)\sigma_2\times\II_{28}$
&$2\omega_1$
&$56$ - dimensional
&$\alpha_1+$
\\
\hline
$\IZ_3$
&$E_6$
&$(1/3){\rm y}\times\II_{q}$
&$\omega_1,\ \omega_2$
&$27,\ \bar{27}$
&$\alpha_1+\alpha_2+$
\\
\hline
$\IZ_2\times\IZ_2$
&Spin$(4n)$
&$\sigma_{\mu\nu},\ (1/2)(1\pm\gamma)\sigma_{\mu\nu}$
&$\omega_1,\omega_{r-1},\omega_r$
&$F_1$ (vector), $S^{\pm}$ (spinor)
&$\alpha_1+2\alpha_2+\dots+2\alpha_{r-2}+\alpha_{r-1}+\alpha_r$
\\
\hline
$\IZ_4$
&Spin$(4n+2)$
&$\2\gamma,\ \smallover1/4+\2(1\pm\gamma)\sigma_{\mu\nu}$
&$\omega_1,\omega_{r-1},\omega_r$
&$F_1$ (vector), $S^{\pm}$ (spinor)
&$\alpha_1+2\alpha_2+\dots+2\alpha_{r-2}+\alpha_{r-1}+\alpha_r$
\\
\hline
$\IZ_n$
&SU$(n)$
&$\smallover{1}/{n}{\rm diag\, }(k,n-k)$
&$\omega_k,\ k=1,\dots,n-1$
&$n-1$ primitive reps. $F_k$
&$\alpha_1+\alpha_2+\dots+\alpha_{n-1}+\alpha_{n-1}$
\\
\hline\\
\end{tabular}
\caption{\it The simply connected simple compact Lie groups with non-trivial centres, their minimal co-weights, expressed as matrices and as primitive weights, the representations characterized by co-weights and the expansions of the highest roots in terms of the primitive roots. Here $\sigma_2$ and
$\sigma_3$ denote Pauli matrices, $\gamma_\mu$ Clifford matrices,
${\rm y}$ the $\SU(3)$ hypercharge diag$(2,-1_,-1)$ and $\gamma=\gamma_1\dots\gamma_{4n}$.}
\label{tableau}
\end{sidewaystable}
\goodbreak
On the other hand, the correspondence $W\sim z$ can be made one-to-one by restricting the
$W$'s to those ones, $\oW$'s (say), for which the geodesics
$\widetilde{\exp}[2\pi i\oW t]\ (0\leq t\leq1)$ are geodesics of \emph{minimal length} from $1$ to $z$ i.e. for which $\tr W^2$
is minimal for each $z\in Z$. (Since the weights $W$ are all of different lengths and are unique up to conjugation, the $\oW$
for each $z\in Z$ will be unique up to conjugation). Such co-weights
$\oW$ are called \emph{minimal vectors} or \emph{minimal co-weights} \cite{4}, and a simple intuitive way to find them
(indeed an alternative way to introduce them) is as follows.

Let $z\in Z$ be a central element in the fundamental representation $F$ of the group and let $f$ be the dimension of $F$. Then by Schur's lemma and the unimodularity of $F$ the elements $z$ must be of the form $z=\exp[2\pi i\lambda]\II_f$, where $\lambda=p/f$ and $p$ is an integer between $0$ and $f$.
(Note that if $F$ is real or pseudo-real, $z$ must be real and therefore equal $\pm1$, a result which explains the abundance
of $Z=\IZ_2$'s in Table 1). It is clear that $z$ is an element of the center of $\SU(f)$ as well as of $K$, and hence one may start by constructing the minimal geodesic from $\II_f$ to $z$ in $\SU(f)$. Let this be $\exp[2\pi it\Sigma],\ 0\leq t\leq1$, where
$\Sigma$ is a generator  in $\SU(f)$. Since $\exp[2\pi i\Sigma]=\big(\exp [2\pi i\lambda]\big)\II_f$, the eigenvalues of $\Sigma$ can only be of the form $\lambda+\ell_k,\ k=1,\dots,f$, where the $\ell_k$ are integers, and hence the geodesic length must be proportional to $\sum_k(\lambda+\ell_k)^2$. It is clear that this length will be smaller for $\ell_k=0$ or $(-1)$ than for any other set of $\ell$'s. But since $\Sigma$ must be traceless, there is (up to conjugation) only one $\Sigma$ for which
$\ell_k=0,-1$, namely
\beq
\Sigma=\frac1f\barr{cc}
p\II_q&0\\[6pt]
0&-q\II_p
\earr\,,
\qquad
\hbox{where}\;p+q=f.
\label{2.8}
\eeq

For $K=\SU(n)$ this is the end of the story, since $n=f$ and hence $\oW=\Sigma$. But remarkably, this is the end of the story also
for the other groups. More precisely, for every group given in Table 1, $\oW$ is an $\SU(f)$ conjugate of $\Sigma$. We do not know of a universal (i.e. group-independent) proof of this result, but it is not difficult to verify it for each class of group in Table 1 separately. For this purpose it is convenient to
characterize $\Sigma$ in a conjugation-independent manner, namely to write
\beq
\left(\Sigma-\frac{p}{f}\right)\left(\Sigma+\frac{q}{f}\right)=0,
\label{2.9}
\eeq
since then one has only to verify that the group in question has a generator satisfying (\ref{2.9}) for a given central element i.e.
given fraction $p/f$. Now for the groups with center $\IZ_2$ and $\IZ_2\times\IZ_2$ this equation reduces to $\Sigma^2=\smallover1/4$ and it is easy to verify that
the generators shown in Table 1 have this property. Similarly for the only group with center $\IZ_3$, namely $E_6$, it can be verified directly that it has a generator of the
form $(y/3)\times\II_q$ and that such a generator satisfies 
(\ref{2.9}) for $p/f=1/3$. The class of groups with center $\IZ_4$, namely ${\rm Spin}(4n+2)$, is perhaps the most interesting. In this case
$\Sigma$ should satisfy the equation
\beq
\Sigma^2=\frac14
\quad\hbox{or}\quad
(\Sigma\pm\frac14)(\Sigma+\frac34)=0
\label{2.10}
\eeq
and one can see that the entries of $\oW$ given in Table 1 satisfy
these equations and are generators by recalling that 
${\rm Spin}(4n+2)$ splits into the direct sum of two inequivalent
spin representations of ${\rm Spin}(4n)$ with generators
$(1\pm\gamma)[\gamma_\mu,\gamma_\nu]/2$ respectively,
where $\gamma=\gamma_1\dots\gamma_{4n}$ is the generalization of
$\gamma_5$ to $4n$ dimensions.

Collecting the results for the different groups together, one sees that in all
cases the $\oW$'s in the fundamental representation are matrices with
\begin{itemize}
\item
(i) only two distinct eigenvalues, 

\item
(ii) unit difference between the eigenvalues.
\end{itemize}

\noindent
Since it can be shown that the converse is true (any such matrix is
an $\oW$) the $\oW$ may actually be characterized by this property. Furthermore, since the adjoint representation occurs in the  tensor product $F\times F^*$, the properties (i) and (ii)
may also be expressed by saying that the $\oW$'s can have only eigenvalues $0$ or $\pm1$ in the adjoint representation, and since the converse is again true, the $\oW$'s may
be characterized by this $0,\pm1$ property also.

In terms of roots $\alpha$, the $0,\pm1$ property (crucial in the stability investigation)
may be expressed by saying that for any positive root $\alpha$
the quantity $\alpha(\oW)$ must be
zero or unity \cite{4,5}: $[E_\alpha,W]=\alpha(W)E_\alpha\,\Rightarrow$
\beq
\alpha(\oW)=0,\pm1.
\label{2.11}
\eeq
If one considers in particular the expansion of the highest root $\theta$ in terms
of the primitive roots $\alpha_i$, $\theta=\sum h_i\alpha_i,\
h_i\geq1$, and applies (\ref{2.11}) to both sides of this
equation, one sees that $\alpha_i(\oW)$ can be non-zero for only one primitive root, $\oalpha_i$ (say), and that the coefficient $\stackrel{\smallcirc}{h}_i$ of $\oalpha_i$ must be unity \cite{4,5}.
This result provides us with a simple, practical method of identifying the $\oW$'s in terms of primitive weights, namely as the duals to those primitive roots for which the coefficient in
the expansion of $\theta$ is unity \cite{4,5}.
This method has been used to obtain the identification given in Table 1.

The $W$-lattice containing the charge lattice, together with the root planes, form the \emph{Bott diagram} \cite{12} of $K$.
Those vectors satisfying the condition (\ref{2.11}) either lie in the
center or belong to the root plane which is the closest to the center. Examples are given in Sec. 9.

%%%%%%%%%%%%%%%%%%%%%%%%%%%%%%%%%%%%%%%%%%%%%%%%%
\section{FINITE ENERGY CONFIGURATIONS AND HIGGS BREAKDOWN}
%%%%%%%%%%%%%%%%%%%%%%%%%%%%%%%%%%%%%%%%%%%%%%%%%

Our starting point is a static, purely magnetic Yang-Mills-Higgs (YMH) system with a simple and compact gauge group $G$, given by the Hamiltonian
\beq
E=\int\frac{1}{2}\Big(\tr \bB^2+\tr(\bD\Phi\bD\Phi)+2V(\Phi)\Big) d^3x\,\,,
\qquad V(\Phi)\geq0;
\label{3.1}
\eeq
where $V(\Phi)$ is a Higgs potential for the  scalar field $\Phi$,
$\bB$ is the Yang-Mills magnetic field and $\bD\Phi$ is the covariant derivative, 
$$B_i=\2\epsilon_{ijk}B_{jk},
\qquad
B_{jk}=\nabla_jA_k-\nabla_kA_j-i[A_j,A_k],
\qquad
D_j\Phi=\nabla_j\Phi-iA_j\Phi,
$$
 where $\bA$ is the gauge potential and
$\bA\Phi$ denotes its action on $\Phi$ in the
representation to which $\Phi$ belongs. For example, if
$\Phi$ is in the adjoint representation, $\bA\Phi$ means
$[\bA,\Phi]$.

In this section we shall \emph{not} require that the fields satisfy the Euler-Lagrange
field equations, but only that they be of finite energy, i.e., such that the integral in
(\ref{3.1}) converges. One reason for this is to emphasize that the most important spontaneous symmetry breakdown, namely that of the Higgs potential, comes from the finite energy and not from the field equations.

We shall consider the three terms in the Hamiltonian (\ref{3.1}) in turn. It will be convenient to use the radial gauge $\bx\cdot\bA=0$.

\goodbreak
%%%%%%%%%%%%%%%%%%%%%%%%%%%%%%%%%%
\kikezd{Pure gauge term $\tr \bB^2$}
%%%%%%%%%%%%%%%%%%%%%%%%%%%%%%%%%%

For sufficiently smooth gauge fields the finite energy condition imposed by this term is evidently
\beq
\bA(\bx)\to\frac{\bA(\Omega)}{r},
\qquad
\bB(\bx)\to\frac{\bb(\Omega)}{r^2}
=b(\Omega)\frac{x^i}{r^3},
%B^a_i(\bx)\to\frac{b^a_i(\Omega)}{r^2}
%=b_i(\Omega)\frac{x^a}{r^3}
\label{3.2}
\eeq
where $\Omega$ denotes the polar angles ($\theta,\varphi)$\footnote{$\bA=(A^a_i)$ is a Lie algebra valued vector potential with $a$ and $i$
  Lie algebra and resp. space indices. 
Similarly, $\bB=(B^a_i)$ and $\bb=(b^a_i)$ are Lie algebra valued
vectors. The last equality in
(\ref{3.2}) decomposes the Lie algebra valued 
magnetic field 
into a Lie algebra-valued scalar $b(\Omega)/r^2$
 times the radial direction, see
\cite{HPAMonop}.}.

Although $\bA(\Omega)$ and $\bb(\Omega)$ must be single-valued on the sphere $\IS^2$, they need not be quantized for (\ref{3.2}) to be satisfied. The situation is analogous to an
Aharonov-Bohm potential in two dimensions, where the gauge field is single-valued but the
magnetic flux need not be quantized~\footnote{Only for the so-called
\emph{vortex system}, in which there exists, in addition to the gauge field, a scalar field $\Phi(\bx)$, which remains finite and \emph{covariantly constant} as $r\to\infty$
does the flux become quantized. The generalization of the vortex case will be seen below.}.

%%%%%%%%%%%%%%%%%%%%%%%%%%%%%%%%%%
\kikezd{Higgs potential $V(\Phi)$}
%%%%%%%%%%%%%%%%%%%%%%%%%%%%%%%%%%

The finite energy condition for this term is evidently
$r^2V(\Phi)\to0$ as $r\to\infty$. A necessary condition for this is that $V\to0$. But $V\geq0$ is assumed to be a Higgs potential i.e. minimizes on a non-trivial group orbit $G/H$. Therefore, at large distances, the Higgs field is not zero, 
but takes its values on the orbit $G/H$ and may depend 
nontrivially on the polar angles $\Omega$:
$\Phi(r,\Omega)\to \Phi(\Omega)$ as $r\to\infty$. Then 
$\Phi(\Omega)$ defines a map of $\IS^2$ into the orbit $G/H$ 
and thus a homotopy class in $\pi_2(G/H)$. Since this class
can not be changed by smooth deformations \cite{8}, the part of finite-energy
configurations splits into \emph{topological sectors},
labelled by $\pi_2(G/H)$.

As well-known, the topological sectors can be labelled also by
certain classes in $\pi_1(H)$. Indeed, on the upper and respectively on the lower hemispheres $N$ and $S$ of $\IS^2$,
$\Phi(\Omega)=g_N(\Omega)\Phi(E)$ in $N$ and
$\Phi(\Omega)=g_S(\Omega)\Phi(E)$ in $S$, where $E$ is
an arbitrary point in the overlap, the ``east pole''.
\beq
h(\varphi)=g_N^{-1}(\varphi)g_S(\varphi)
\label{3.3}
\eeq
(where $\phi$ is the polar angle on the equator of $\IS^2$)
is a loop in $H$ which represents the topological sector. (\ref{3.3}) is contractible in $G$ \cite{2,8}.

For any compact and connected Lie group $H$, $\pi_1(H)$
is of the form
\beq
\pi_1(H)=\IZ^p\oplus\IT,
\label{3.4}
\eeq
where $p$ is the dimension of the center $Z$ of $H$
and $\IT$ is a finite Abelian group \cite{22}. In fact,
$\IT$ is isomorphic to $\pi_1(K)$, where $K$ is the compact and semisimple subgroup of $H$ generated by
$\gk=[\gh,\gh]$.

The free part $\IZ^p$ provides us with $p$ integer
``quantum'' numbers $m_1,\dots, m_p$. They can be calculated 
as surface integrals as follows. To a physical Higgs field
$\Phi(\Omega)$ in \emph{any} representation and to each vector $\Psi$ from the center of the Lie algebra $\gh$, we can associate a new, adjoint ``Higgs'' field $\Psi(\Omega)$ defined by
$\Psi(\Omega)=g(\Omega)\Psi g^{-1}(\Omega)$, where $g(\Omega)$
is  any of those ``lifts'' in (\ref{3.3}). $\Psi(\Omega)$ is
well-defined, because $\Psi$ belongs to the center.
In particular, the projections of the charge lattice
$\Gamma_Q$ into the center is a $p$-dimensional lattice there, generated over the integers by $p$ vectors 
$\Psi_1,\dots\Psi_p$. The above construction associates then
and adjoint ``Higgs'' field $\Psi_i(\Omega)$ to each 
generator $\Psi_i$, and the quantum numbers $m_k$ are calculated according to
\beq
m_k=\frac{1}{2\pi}\int d\Omega_{ij}\tr\left(\Psi_k(\Omega)
\big[\p_i\Psi_k(\Omega),\p_j\Psi_k(\Omega)\big]\right).
\label{3.5}
\eeq
(The finite part of $\pi_1(H)$ has no similar expression.)

The physically most relevant case is when the homotopy group
$\pi_1(H)$ is described by a single integer quantum number $m$. This happens when the Lie algebra $\gh$ of $H$ has a $1$-dimensional center generated by a single vector $\Psi$ and the
semisimple subgroup $K$ is simply connected. This happens in particular when the Higgs field $\Phi(\Omega)$ belongs to the adjoint representation of a classical group $G$, and the Higgs
potential $V(\Phi)$ is quartic: this is the content of the Michel conjecture \cite{23}. In fact,
for the adjoint representation of a classical group, $\Phi$ itself generates the center and is parallel to one of the
primitive (but not necessarily minimal) $W_j$'s.
More generally, if the Higgs field is in the adjoint representation, $\IT$ is always trivial so that $\pi_1(H)=\IZ^p$.

The homotopy classification is not merely convenient, 
but is \emph{mandatory} in the
sense that the classes are separated by infinite
energy barriers. Thus, while an interpolated field of the form 
$\Phi^\tau=\tau\Phi_1+(1-\tau)\Phi_2,\ 0\leq\tau\leq1$ 
between two finite-energy configurations 
$\Phi_1$ and $\Phi_2$ is perfectly smooth if 
$\Phi_1$ and $\Phi_2$ are smooth, it does not satisfy the finite-energy condition 
$r^2V(\Phi^\tau)\to0$, or even $V(\Phi^\tau)\to0$,
as $r\to\infty$, for general $\tau$.

Note that since not only $V\to0$ but $r^3V\to0$ one has, using the notation $\eta=\Phi(r,\Omega)-\Phi(\Omega)$,
\beq
r^3M_{\alpha\beta}\eta_\alpha\eta_\beta\to0
\qquad
\hbox{where}
\qquad
M_{\alpha\beta}=\frac{\p^2V}{\p\Phi_\alpha\p\Phi_\beta}\Big|_{r=\infty}
\label{3.6}
\eeq
and hence for generic potentials (i.e. those for which the only zeros of the `mass matrix'  ${\p^2V}/{\p\Phi^2}$
 at $V=V_{min}$ are the Goldstone zeros) the physical part of 
$\eta$ falls off faster than $r^{-1}$ as $r\to\infty$ and one gets
$\Phi(\bx)\to\Phi(\Omega)+\eta(r,\Omega)$, where
$r\eta(r,\Omega)\to0$ as $r\to\infty$. A notable exception
to this observation is the Bogomolny-Prasad-Sommerfield (BPS)
case $V=0$, for which the Bogomolny condition $\bB=\bD\Phi$
implies \cite{14} that 
\beq
\Phi(\bx)\to\Phi(\Omega)+\frac{b(\Omega)}{r}+
\Ort(1/r^2)
\qquad\hbox{as}\qquad
r\to\infty.
\label{3.7}
\eeq

%%%%%%%%%%%%%%%%%%%%%%%%%%%%%%%%%%%%%
\kikezd{The cross-term $(\bD\Phi)^2$}
%%%%%%%%%%%%%%%%%%%%%%%%%%%%%%%%%%%%%

This final term involves both $\Phi$ and $\bA$ and it hence provides the connection between the Higgs field $\Phi(\Omega)$
 and the
gauge field $\bb(\Omega)$ and thus puts a topological constraint
on the gauge field. As might be expected from the vortex analogy,
this constraint may be expressed as a quantization condition as follows: the finite energy
condition is easily seen to be $r^2(\bD\Phi)^2\to0$ and thus
$\Phi(\Omega)$ and $\Psi(\Omega)$ are hence both covariantly
constant on $\IS^2$,
\beq
\bd\Phi\equiv\bpar\Phi-i\bA(\Omega)\Phi(\Omega)=0,
\qquad
\bd\Psi\equiv\bpar\Psi-i[\bA(\Omega),\Phi(\Omega)]=0,
\label{3.8}
\eeq
where $\bpar=r\bnabla$.

The topological quantum numbers $m_k$ can be expressed in this case as
\beq
m_k=\frac{1}{2\pi}\int d\Omega\tr(\Psi_kb),
\qquad
k=1,\dots,p.
\label{3.9}
\eeq
Equation (\ref{3.9}) is the generalization of the vortex quantization condition mentioned
earlier and it shows that in general it is not the gauge field
$b$ itself, but only its \textit{projection onto the center}
that is quantized. Note that the quantization of $\int\tr(\Psi_kb)$
is again \emph{mandatory}, since the value of $\tr(\Psi_kb)$
cannot be changed without violating at least one of the finite-energy conditions $r^2V\to0$ or $r^3(\bD\Phi)^2\to0$ and
thus passing through an infinite energy barrier.

Notice that the value of (\ref{3.8}) is actually independent of the choice of the Yang-Mills potential $\bA$ as long as $\Phi$ is covariantly constant \cite{21}.

If $\bD\Phi=0$, a loop representing the homotopy sector can be found by parallel transport \cite{3,8}. Indeed, let us cover $\IS^2$ by a $1$-parameter family of loops $\gamma_{\varphi}(\theta)$,
e.g., by choosing $\gamma_{\varphi}$ to start from the north pole, follow the meridian
at angle $\varphi=0$ down to the south pole and return then to the north pole along the meridian at angle $\varphi$.
The loop
\beq
h^A(\varphi)={\cal P}\left(\exp\oint_{\gamma_{\varphi}}\bA\right)
\label{3.10}
\eeq
then represents  the topological sector.

 Other choices of the $1$-parameter family of paths
$\gamma_{\varphi}$ would lead to homotopic loops $h^A$.

%%%%%%%%%%%%%%%%%%%%%%%%%%%%%%%%%%%%%%%%%%%%%%%%%
\section{FINITE ENERGY SOLUTIONS OF THE FIELD EQUATIONS}
%%%%%%%%%%%%%%%%%%%%%%%%%%%%%%%%%%%%%%%%%%%%%%%%%

The only condition imposed on the YMH configurations $(\bA,\Phi)$ up to this point is that the energy be finite. But it is obviously of interest to consider the special case of finite energy configurations that are also solutions of the YMH field equations,
\beq
\bD^2\Phi=\frac{\p V}{\p \Phi}
\quad\hbox{and}\quad
\bD\times\bB=(\Phi,\btau\,\bD\Phi)
\label{4.1}
\eeq
where $\btau$ denotes the generators of the Lie algebra in the
representation to which the Higgs field $\Phi$ belongs.

Finite energy \emph{solutions} may be classified using data
referring to the field $b(\Omega)$ alone. For this it is sufficient to consider the field equations (\ref{4.1}) for
large $r$, in which case they reduce to
\beq
\bigtriangleup\eta_\alpha=\left(\frac{\p^2V}{\p\phi_\alpha\p\Phi_\beta}\right)\eta_\beta
\quad\hbox{and}\quad
\bd\times\bb=0
\label{4.2}
\eeq
in the generic  case (and to $\bigtriangleup\eta=0$ and
$\bd\times\bb=0$ in the Bogomolny case). The first equation shows that, for solutions of (\ref{4.1}), the generic finite-energy condition $\eta\to0$ is sharpened to an exponential fall-off of $\eta$. (The BPS case escapes because
$\bD^2\eta=0$ is consistent with $\eta=b(\Omega)/r$.)

Since $\Phi(\Omega)$ and $b(\Omega)$ are the only components of the
field configuration that survive in the asymptotic region, within each topological sector defined by $\Phi(\Omega)$,
the only possible asymptotic classification of the configurations is according to $b(\Omega)$. The conditions
satisfied by $b(\Omega)$ are then contained in the second equation in
(\ref{4.2}), which may be written as
\beq
\bd b\equiv \p b-i[\bA(\Omega),b]=0.
\label{4.3}
\eeq
This equation shows that $b(\Omega)$ is covariantly constant and thus lies on an $H$-orbit. Therefore $b(\Omega)=
h_N(\Omega)Qh_N^{-1}(\Omega)$ in $N$ and 
$b(\Omega)=
h_S(\Omega)Qh_S^{-1}(\Omega)$ in $S$, where 
$Q=b(E)$ is in $\gh$. Plainly, $Q$ is unique up to 
global gauge rotations, and there is thus no loss of generality in choosing it in a given Cartan algebra. In the singular gauge where $b(\Omega)=Q$, the loop (\ref{3.10}) is
simply
\beq
h(\varphi)=\exp[2iQ\varphi],
\qquad
0\leq\varphi\leq2\pi
\label{4.4}
\eeq
and the periodicity of $\varphi$ provides us with the
\emph{quantization condition} 
$$
\exp4\pi iQ=1
$$
so that \emph{$2Q$ is a charge}. Conversely, any quantized $Q$ defines an asymptotic solution, namely 
\beq
\bA=\bA^DQ
\qquad\hbox{i.e.}\qquad
A_\theta=0,\quad
A_\varphi=\pm(1\mp\cos\theta)Q,
\eeq
in the Dirac gauge, so that $b=Q$ and (\ref{4.4}) is the transition function. Solutions can thus be classified by \emph{charges of $H$.}

According to (\ref{3.9}), for solutions of the field equations
the expression for the ``Higgs'' quantum numbers $m_k$
reduces to
\beq
m_k=\frac{2\tr(Q\Phi_k)}{\tr(\Psi_k^2)},
\qquad
k=1,\dots,p.
\label{4.5}
\eeq
Let us now consider a charge $Q$ and denote its topological sector by
$m$. let us decompose $Q$ into central and semisimple parts
$Q_{||}$ and $Q_{\perp}$, respectively. By (\ref{4.5}),
$$
2Q_{||}=\sum_km_k\Psi_k.
$$
Observe that
\beq
z=\exp[4\pi iQ_{||}]=\exp[-4\pi iQ_{\perp}]
\label{4.6}
\eeq
lies simultaneously in $Z(H)_0$ (the connected component 
of the center of $H$) and in the semisimple subgroup $K$, and thus also in $Z(K)$, the center of
$K$. Let us decompose $\gk=[\gh,\gh]$ into simple
factors, 
$$
\gk=\gk_1\oplus\dots\oplus\gk_s,
$$
and denote by $\widetilde{K}_j$ the simple and simply connected group, whose algebra is $\gk_j$. As explained in
Sec. 2, $K$ is of the form $\widetilde{K}/C$, where
$C=C_1\times\dots\times C_s$ is a subgroup of the
center $Z=Z(\widetilde{K})$ of $\wK=\big[\wK_1\times\dots\times\wK_s\big]$, $C_j$ being a subgroup of $Z(\wK_j)$.

The situation is particularly simple when $K$ is simply connected,
$K=\wK$, when the central part $Q_{||}$ contains all
topological information. Indeed, $z$ is uniquely
written in this case as
\beq
z=z_1\dots z_s,
\qquad\hbox{where}\qquad
z_j\in Z(\wK_j).
\label{4.7}
\eeq
However, as emphasized in Sec. 2, the central elements of a simple and simply connected group are in one-to-one correspondence with the minimal
$\oW$'s and thus, for each $z$ in the center, there exists a unique set of $\oW_j$'s (where $\oW_j$ is
either zero or a minimal vector of $\gk_j$) such that
\beq
z=\big(\exp[-2\pi i\oW_1]\big)\dots
\big(\exp[-2\pi i\oW_s]\big)=
\exp\big[-2\pi i\sum_{k=1}^s\oW_k\big]
=
\exp\big[-2\pi i\oW^{(m)}\big].
\label{4.8}
\eeq
$\oW^{(m)}$ depends only on the sector (and not on $Q$ itself), because all charges of a sector have
the same $Q_{||}$. Hence the \emph{entire sector}
can be characterized by giving
\beq
2\oQ^{(m)}=\sum_km_k\Psi_k+\oW^{(m)}
\label{4.9}
\eeq

By (\ref{4.8}) $2\oQ^{(m)}$ is again a charge, 
$\exp[4\pi\oQ^{(m)}]=1$, and it obviously belongs
to the sector $m$. Furthermore, 
$$
\exp[4\pi i(Q-\oQ)]=\exp[4\pi iQ]\exp[-4\pi i\oQ]=1$$
shows that $2Q'=2(Q-\oQ)$ is in the charge lattice of $K$.

The situation is slightly more complicated if $K$ is
non-simply- connected, so that the semisimple part also contributes to the topology. Since $C$ is now non-trivial,
the expansion (\ref{4.7}) is not unique, and $z_j$ can be replaced rather by
$z_j^*=z_jc_j$, where $c_j$
belongs to the subgroup $C_j$ of $Z(\wK_j)$. 
But $z^*_j$ is just another element of
$Z(\wK_j)$, so it is uniquely $z^*_j=\oW_j^*$
for some minimal $\oW_j^*$ of the simple factor $\wK_j$.
Equation (\ref{4.8}), with all $\oW_j$'s replaced by the $\oW_j^*$'s, is still valid, so that (\ref{4.9}) is a charge also now. However, since $\pi_1(K)=C=C_1\times\dots\times C_s$, those loops generated by $Q$ and
$Q^*$ belong now to different topological sectors.

We conclude that a topological sector contains a unique charge $Q$ of the form (\ref{4.9}) also in this case, and that, in full generality, any other monopole charge
is uniquely of the form
\beq
Q=\oQ+Q'=\oQ+\2\sum_i^rn_iQ_i,
\label{4.10}
\eeq
where the $n_i$ are integers, and the $Q_i,\,i=1,\dots r$
are the primitive charges of $K$. (Obviously, the $Q_i$ are sums of primitive charges taken for the simple factors $K_j$).
The integers $n_i$ could be regarded as secondary quantum
numbers which supplement the Higgs charge $m$.

In Sec. 5, we shall show that $\oQ^{(m)}$ is the
\emph{unique stable monopole} in the sector $m$.
The situation is conveniently illustrated on the Bott
diagram, see Section 9. 

The classification of finite energy solutions according to the secondary quantum numbers or, equivalently the
matrix-valued charge $Q$ is convenient and illuminating,
but in contrast to the classification of finite energy configurations according to the Higgs charge $m$, it is
\emph{not} mandatory, in the sense that (for fixed $m$) the different charges $Q$ are separated only by \emph{finite} energy barriers, see Sec. 8.

%%%%%%%%%%%%%%%%%%%%%%%%%%%%%%%%%%%%%%%%%%%%%%%%%
\section{UNSTABLE SOLUTIONS: REDUCTION FROM $\IR^3$ 
TO $\IS^2$}
%%%%%%%%%%%%%%%%%%%%%%%%%%%%%%%%%%%%%%%%%%%%%%%%%

Now we wish to show that those monopoles for which
$Q'\neq0$ are unstable. More precisely, we show that for a restricted class of variations the stability problem reduces to a corresponding Yang-Mills problem on
$\IS^2$. This allows us to prove that with respect to our variations there are
\beq
\nu=2\sum_{q<0}\left(2|q|-1\right)
\label{5.1}
\eeq
independent negative modes.

To prove our statement, let us first introduce the
notation
\beqa
\big(\ba\times\bb\big)_i&=&\varepsilon_{ijk}a_jb_k,
\label{5.2}
\\[6pt]
\big[\ba\times\bb\big]&=&\ba\times\bb-\bb\times\ba
\quad\hbox{i.e.}\quad
\left(\big[\ba\times\bb\big]\right)_i=
\varepsilon_{ijk}\big[a_j,b_k\big].
\nn
\eeqa
Note that $\ba\times\ba$ may be different from zero if 
$\gh$ is non-Abelian.

For $\gh$-valued variations of the \emph{gauge potentials alone}, $\delta\Phi=0,\ \delta\bA=\ba\in\gh$ say,
the variations of the gauge field and covariant derivative are easily seen to be 
$\delta\bB=\bD\times\ba$,  $\delta^2\bB=-i[\ba\times\ba]$
and $\delta(\bD\Phi)=-i\ba\Phi$. All higher-order variations $\delta^3\bB$ etc. are zero.

For the energy functional (\ref{3.1}) the first variation is zero, since $(\bA,\Phi)$ is a solution of the field equations, and the higher order variations are
\beq\begin{array}{lll}
\delta^2E&=&\displaystyle\int d^3x\big\{
\tr(\bD\times\ba)^2-i\tr(\bB[\ba\times\ba])
-\tr(\ba\Phi)^2\big\}
\\[8pt]
\delta^3E&=&-3i\displaystyle\int d^3x\tr\big\{
(\bD\times\ba)(\ba\times\ba)\big\},
\\[8pt]
\delta^4E&=&-3\displaystyle\int d^3x\tr(\ba\times\ba)^2,
\end{array}
\label{5.3}
\eeq
all higher-order variations being zero. We shall assume that all variations are square-integrable, 
$$
(\ba,\ba)=\displaystyle\int drd\Omega\tr(\ba)^2<\infty.
$$

There are some general points worth noting. First, since $\delta\Phi=0$,
the only terms in (\ref{5.2}) involves the Higgs field is
$\tr(\ba\Phi)^2$ and since $\ba$ must be in the little
group of $\Phi(\Omega)=\lim_{r\to\infty}\Phi(r,\Omega)$,
for $V\neq0$ ('t Hooft-Polyakov case) this term vanishes asymptotically. Thus, if we only consider asymptotic variations \cite{1,2} i.e. such that $\ba(r,\Omega)=0$
for $r\leq R$ where $R$ is `sufficiently large', and in practice this will mean $R$ large enough for the asymptotic form (\ref{3.2}) of the fields to be valid,
we can then drop the Higgs terms in (\ref{5.3}) and consider the pure Yang-Mills variations
\beq
\delta^2E=\int d^3x\big\{
\tr(\bD\times\ba)^2-i\tr(\bB[\ba\times\ba])
\big\}\ .
\label{5.4}
\eeq

Second, the only term in (\ref{5.4}) that involves radial
derivatives is the $(\nabla_r\ba)^2$ term in
$(\bD\times\ba)^2$ and this contribution may be written as
\beq
\delta^2E_r=\int d^3x\tr(\nabla_r\ba)^2=
\int drd\Omega\tr(r\nabla_r\ba)^2=
\int drd\Omega\tr\big\{(d\ba)^2+\smallover1/4\ba^2\big\}
=m^2(\ba,\ba),
\label{5.5}
\eeq
where $d$ is the symmetrical dilatation operator 
$\2\{r,\nabla_r\}$, $\delta^2$ is its average value,
and $m^2=\smallover1/4+\delta^2$. It can be shown (see Appendix) that the infimum of $\delta^2$ is
$0$, and thus, although $\delta^2E_r$ is not negligible because of the lower bound $\smallover1/4$, it can be
reduced to this lower bound, and $\delta^2E_r$ can be regarded as a mass term. Thus the variations (\ref{5.4})
are essentially variations on the $2$-sphere $\IS^2$, for each value of $r$.

Finally, it should be noted that some of the variations, namely $\ba=\bD\chi+{\rm O}(\chi^2)$ where $\chi$ is
any scalar, are simply gauge transformations of the background field $\bA$ and lead to zero-energy variations. In particular, it is easy to verify that, because $\bA$ satisfies the field equations, the second
variation $\delta^2E$ is zero for the infinitesimal variations $\delta\bA=\bD\chi$, and for this reason it is convenient to define the `physical' variations
$\ba$ as those which are orthogonal to the $\bD\chi$.
Since $\chi$ is arbitrary, one has 
\beq
\displaystyle\int d^3x\tr(\ba\,\bD\chi)=\displaystyle\int d^3x\tr(\bD\ba\,\chi)=0\,\quad\Rightarrow\,\quad
\bD\cdot\ba=0,
\label{5.6}
\eeq
from which one sees that the physical variations may also be characterized as those which are divergence-free. As a consequence of the gauge condition $A_r=0$ our variations satisfy also $a_r=0$.

It will be convenient to write (\ref{5.4}) in the form
\beqa
\delta^2E&=&\delta^2E_1+\delta^2E_2
\label{5.7}
\\[8pt]
&&=\displaystyle\int d^3x\tr\left\{(\bD\times\ba)^2
+(\bD\cdot\ba)^2\right\}
+\displaystyle\int d^3x\tr\left\{-i\bB[\ba\times\ba]-(\bD\cdot\ba)^2\right\},
\nn
\eeqa
bearing in mind that $\bD\cdot\ba$ is unphysical and may be gauged to zero.

Let us first consider $\delta^2E_1$. From the identity
\beq
\big(\bD\times(\bD\times\ba)\big)_j=-\bD^2a_j+
D_j(\bD\cdot\ba)-i\big([\bB\times\ba]\big)_j
\label{5.8}
\eeq
using $\bb=\lim_{r\to\infty}r^2\bB$, one sees that
\beqa
\delta^2E_1&=&\int d^3x\tr\left\{(-\bD^2\ba-i[\bB\times\ba])\ba\right\}=\delta^2E_r+\delta^2E_{\Omega}
\nn
\\[8pt]
&=&\delta^2E_r+\int d^3x r^{-2}\tr\left\{\big(\bL^2\ba-
i[\bb\times\ba]\big)\ba\right\},
\label{5.9}
\eeqa
where $\bL=-i\bx\times\bD$ is the orbital angular momentum.

It is well-known that the components of $\bL$ do not satisfy the angular momentum relation, $[L_i,L_j]=i\varepsilon_{ijk}\big(L_k+x_k(\bx\cdot\bB)\big)
\neq i\varepsilon_{ijk}L_k$, but that for spherically
symmetric, and hence for asymptotic, field, the quantity $\bM$ obtained by subtracting $\bx(\bx\cdot\bB)=
\bb=\frac{\bx}{r}b$ [cf. (\ref{3.2})] from $\bL$ does satisfy such an algebra i.e.
\beq
\big[M_i,M_j\big]=\varepsilon_{ijk}M_k
\qquad
\hbox{where}\qquad
M_i=L_i-\frac{x_i}{r}\,b .
\label{5.10}
\eeq
$\bM$ in (\ref{5.10}) is the angular momentum for a spinless particle. (Remember that for a particle $\psi$ in the adjoint representation for example, $b\psi$ means 
$[b,\psi]$.)

For arbitrary variations $\ba$ the spectrum of $\delta^2E_1$ could be obtained  directly from the conventional $\so(3)$ spectrum of $\bM^2$. But
it is more convenient to use instead the spin-$1$ angular momentum operator
\beq
\bJ=\bM+\bS
=-i\bx\times\bD-\bb+\bS
\label{5.11}
\eeq
where $\bS$ is the $3\times3$ spin matrix
 $\big(\bS_i\big)_{jk}=i\varepsilon_{ijk}$.
$\bS$ satisfies the relations
\beqa
[S_i,S_j]=i\varepsilon_{ijk}S_k,
\nn\\[6pt]
(\bb\cdot\bS)\ba=(b_iS_i)\ba=i[\bb\times\ba],
\label{5.12}
\\[6pt]
\bS^2=S_iS_i=-2.
\nn
\eeqa
Using the gauge conditions $\bD\cdot\ba$ and $\bx\cdot\ba=0$, we see that
\beq
(\bx\times\bD)\times\ba=\bx(\bD\cdot\ba)-x_i\bD a_i=
-x_i\bD a_i=\ba-\bD(\bx\cdot\ba)=\ba,
\label{5.13}
\eeq
i.e., $\bL\cdot\bS=1$. Since $\bx$ and $\bL$ and thus $\bb$ and $\bL$ are orthogonal, this implies,
\beq
\bJ^2\ba=\bL^2\ba+[\bb\times[\bb\times\ba]]-
2i[\bb\times\ba].
\label{5.14}
\eeq
This leads finally to re-writing $\delta^2E_\Omega$ as
\beq
\delta^2E_\Omega=\int drd\Omega\tr\left\{\big(\bJ^2\ba-[\bb\times[\bb\times\ba]]+i[\bb\times\ba]\big)a\right\}.
\label{5.15}
\eeq
It is convenient to decompose the variation $\ba$ into eigenmodes of $[\bb\times\,\cdot\,]$ i.e. to write
\beq
[\bb\times\ba]=iq\,\ba
\quad\hbox{i.e.}\quad
\varepsilon_{ijk}[b_j,a_k]=[b_{ki},a_k]=iq\,a_i,
\label{5.16}
\eeq
where the $q$'s are the eigenvalues. This is possible and the $q$'s will be real because the $b_{ij}$ is skew-symmetric in the Lie algebra as well as in the vector space, and indeed, because of this, the $q$'s come in pairs
of opposite sign and multiplicity two i.e. in
quadruplets $(q,q,-q,-q)$, see the next Section.

On each $q$-sector $\delta^2E_1$ will be
\beq
\delta^2E_1=m^2(\ba,\ba)+\int drd\Omega\tr
\left\{\big(\{\bJ^2-q(q+1)\}\ba\big)\ba\right\}.
\label{5.17}
\eeq
But $\bJ$ is the Casimir of the angular momentum algebra generated by $\bJ$,
so $\bJ^2=j(j+1)$, where $j$ is integer or half-integer,
according as $q$ is integer or half-integer,
because $q$ is the only non-orbital contribution to $\bJ$. Now since $\delta^2E_1$ is manifestly positive,
we must have
\beq
m^2+\big\{\bJ^2-q(q+1)\big\}=
\big\{\smallover1/4+\delta^2\big\}
+\big\{j(j+1)-q(q+1)\big)\geq0
\label{5.18}
\eeq
and since $\delta^2$ is arbitrarily small, we see that $j\geq|q|-1$. Note that $j\geq|q|-1$ follows
from the manifest positivity of $\delta E_1$.

Equation (\ref{5.18}) implies that the possible values of $j$ are $|q|-1, |q|, |q|+1,\dots $.
In particular, the value of $j=|q|-1$ can occur only for $q\leq-1$, and as it corresponds
to the case when $\delta^2E_1$ is purely radial, it implies 
 that $\bD\cdot\ba=0$, so that the states corresponding to it are physical. Thus we can write
$$
\delta^2E_1=m^2(\ba,\ba)
\quad\hbox{for}\quad
j=|q|-1,\quad q\leq-1
$$
and
\beq
\delta^2E_1=\big\{m^2 +(j-q)(j+q+1)\big\}(\ba,\ba)
\quad\hbox{for}\quad
j\geq|q|\ .
\label{5.19}
\eeq

Let us now consider $\delta^2E_2$. Since $\bD\cdot\ba$ is zero on the physical states,
\beqa
\delta^2E_2&=&(-i)\int d^3x\tr\big(\bB\big[\ba\times\ba\big]\big)=
\int d^3x\tr\left\{-i\big[\bB\times\ba\big]\ba\right\}
\nn
\\[6pt]
&=&\int d^3x\, r^{-2}\tr\left\{-i\big[\bb\times\ba\big]\ba\right\}
=q\int d^3x\,r^{-2}\tr\big(\ba,\ba\big)=q\big(\ba,\ba\big).
\label{5.20}
\eeqa
From the positivity of $\delta^2E_1$ we then see that the Hessian 
$\delta^2E$ will be positive unless $q$ is negative. Furthermore, when $q$ is negative, (\ref{5.19}) becomes
\beq
\delta^2E_1=\big\{m^2+(j+|q|)(j-|q|+1\big\}
\big(\ba,\ba\big)\geq2|q|(\ba,\ba)
\label{5.21}
\eeq
so, for $j\geq|q|$, the restriction of $\delta^2E_1$ to
the physical states will dominate $\delta^2E_2$ 
and the Hessian will again be positive. It follows that the only possibility for getting negative
modes is when $q\leq-1$ and $j=|q|-1$, in which case
\beq
\delta^2E=\big(m^2-|q|\big)\big(\ba,\ba\big)<0.
\label{5.22}
\eeq
Thus finally we have the result that the monopole is \emph{unstable} if, and only if, there is an eigenvalue $q$ such that $|q|\geq1$. The opposite condition
\beq
|q|\leq\2
\label{5.23}
\eeq
is, of course, just the \textit{Brand-Neri stability condition} \cite{1,2,4}. From the discussion of
Sec. 4. we know however that $|q|\leq\2$ if and only if
$Q=\oQ$ [cf. (\ref{4.8})]: $\oQ$ is the
\emph{unique stable charge} of the topological sector.

Note that since in the case $j=|q|-1$ the first term on the right-hand side of (\ref{5.7}) vanishes, the variation actually satisfies the \emph{first-order} equations
\beq
\bd\times\ba=0,
\qquad
\bd\cdot\ba=0
\label{5.24}
\eeq
where $\bd=r\bD$. In particular, they are true physical modes These modes form furthermore a 
$$
2j+1=2|q|-1
$$
dimensional multiplet of the $\bJ$ algebra. We shall see in the next Section that for each $|q|$ there is one and only one such multiplet. Taking into
account the fact that the eigenvalues come in pairs, this proves the index
formula (\ref{5.1}).

Notice that our approach above shows strong similarities to the stability investigations in $\sigma$-models \cite{26}.

The simplest way of counting the number of instabilities for $j\geq |q|$ is to use the Bott
\cite{12} diagram (see the examples of Sec. 9):
(\ref{5.1}) is \emph{twice the number of times
the straight line drawn from the origin to $2Q$
intersects the root planes}.

For \emph{BPS} monopoles the above arguments break down: 
due to the
$b/r$ term in the expansion of the Higgs field, the second variation picks up an extra term $-\tr\big([\ba,b]\big)^2=q^2$ which just \emph{cancels} the $-q^2$ in Eq. (\ref{5.17}). The
total Hessian is thus manifestly positive,
\beq
\delta^2E=\delta^2E_1+\delta^2E_2=
\big((m^2+\bJ^2)\ba,\ba\big)>0.
\label{5.25}
\eeq
We conclude that BPS monopoles are \emph{stable} under variations
of the gauge field alone, even if their charge is \emph{not}
of the form (\ref{4.9}).

%%%%%%%%%%%%%%%%%%%%%%%%%%%%%%%%%%%%%%%%%%%%%%%%%
\section{NEGATIVE MODES}
%%%%%%%%%%%%%%%%%%%%%%%%%%%%%%%%%%%%%%%%%%%%%%%%%

It is convenient to use the stereographic coordinate $z$ on $\IS^2$, 
$z=x+iy=e^{i\varphi}\tan\theta/2$. In stereographic coordinates the background gauge-potential and field strength become
\beq\begin{array}{lllllllllc}
A_x&=&-2Q\displaystyle\frac{y}{\varrho},%&&
&A_y&=&\displaystyle\frac{2Qx}{\varrho},%&&
&b_{xy}&=&-b_{yx}= \displaystyle\frac{4Q}{\varrho^2},
\\[12pt]
A_z&=%&\2(A_x-iA_y)&=
&-iQ \displaystyle\frac{\bz}{\varrho},
&A_{\bz}&=%&\2(A_x-iA_y)&=
&iQ \displaystyle\frac{z}{\varrho},
\qquad
&b_{z\bz}&=&2i \displaystyle\frac{Q}{\varrho^2},
\end{array}\ .
\label{6.1}
\eeq
where $\varrho=1+z\bz$, and
we have treated $z$ and its conjugate $\bz$ as independent variables.
Set $\p=\p_{z},\ \bp=\p_{\bz}$, and  let us define
\beq\begin{array}{llllll}
d_z&=&\p-iA_z=\p-Q\displaystyle\frac{\bz}{\varrho}=
\2(d_x-id_y),
\qquad
&a_z&=&\2(a_x-ia_y)
\\[12pt]
d_{\bz}&=&\bp-iA_{\bz}=
\bp+Q\displaystyle\frac{z}{\varrho}=
\2(d_x+id_y),
\qquad
&a_{\bz}&=&\2(a_x+ia_y)\end{array}
\label{6.2}
\eeq

In complex coordinates the eigenspace-equations (\ref{5.16}) become
\beq\barr{cc}
Q&0
\\
0&-Q
\earr
\barr{c}a_{\bz}\\
a_z\earr
=q\barr{c}a_{\bz}\\
a_z\earr .
\label{6.3}
\eeq
(Remember that $Q$ acts on $a_\alpha$ by conjugation). The
general solution of (\ref{6.3}) is
\beq
\barr{c}a_{\bz}\\
a_z\earr=
f\barr{c} E_\alpha\\ 0\earr
+
g\barr{c} 0\\ E_{-\alpha}\earr,
\label{6.4a}
\eeq
with eigenvalue $q=\alpha(Q)$, where
$f$ and $g$ are arbitrary functions 
of $z$ and $\bz$. Similarly,
\beq
\barr{c}a_{\bz}\\
a_z\earr=
h\barr{c} E_{-\alpha}\\ 0\earr
+
k\barr{c} 0\\ E_\alpha\earr
\label{6.4b}
\eeq 
(where $h$ and $k$ are again arbitrary) are also eigenfunctions 
with eigenvalue $q=-\alpha(Q)$. Equations 
(\ref{6.4a})-(\ref{6.4b}) show that the eigenvalues come in pairs, as stated earlier.

Let us first consider the negative modes, for which we have already seen that $q$ must be negative. As discussed in Sec. 5, for each fixed $q=\alpha(Q)<0$, the negative modes are solutions to the two coupled equations in (\ref{5.23}). Multiplying one of these equations by $i$ and adding and
subtracting the result one sees that (\ref{5.23}) are equivalent to
\beq
d_zf=(\varrho\,\p+|q|\bz)f=0
\quad\hbox{and}\quad
d_{\bz}g=(\varrho\,\bp+|q|z)g=0
\label{6.5}
\eeq
($q=-|q|$ because $q$ is negative). One sees that $f$ and $g$ must be
of the form
\beq
f(z,\bz)=\varrho^{-|q|}\Phi(\bz),
\qquad
g(z,\bz)=\varrho^{-|q|}\Psi(z),
\label{6.6}
\eeq
where $\Phi(\bz)$ and $\Psi(z)$ are arbitrary antiholomorphic (respectively holomorphic)
functions. They can be therefore expanded in power series,
$\Phi(\bz)=\sum c_n{\bz}^n$, and $\Psi(z)=\sum d_mz^m$. But they
are also square integrable functions. Now, since
in stereographic coordinates, the inner product for two vector fields is
\beq
(\ba,\bb)=\int dzd\bz\sqrt{g}g^{\alpha\beta}a_\alpha\bar{b}_\beta=\int dzd\bz\, a_\alpha\bar{b}_\alpha=
\int dzd\bz\,(a_zb_{\bz}+a_{\bz}b_z),
\label{6.7}
\eeq
because $\sqrt{g}g^{\alpha\beta}$ is unity, one sees that $\Phi(\bz)$ will be square integrable if, and only if,
$c_n,\ d_m=0$ except for $n,m=0,1,\dots,2|q|-2$. Thus the negative modes are linear combinations of the $2(2|q|-1)$ variations
\beq
\barr{c}a_{\bz}\\ 0\earr=
\frac{{\bz}^n}{(1+z\bz)^{|q|}}
\barr{c} E_{\alpha}\\ 0\earr
\quad\hbox{and}\quad
\barr{c}0\\ a_{z}\earr=\frac{{z}^m}{(1+z\bz)^{|q|}}
\barr{c} 0\\ E_{-\alpha}\earr,
\label{6.8}
\eeq
$n,m=0,1,\dots,2|q|-2$. These are the $2(2|q|-1)$ multiplets of negative modes referred to in Sec. 5. In polar coordinates
the variations (\ref{6.8}) are also expressed as
\beq\begin{array}{lll}
a_\theta&=&\2e^{\mp i(k+1)\varphi}\big(\sin\smallover{\theta}/{2}\big)^{k}\big(\cos\smallover{\theta}/{2}\big)^{
2|q|-2-k}E_{\mp\alpha},
\\[12pt]
a_\varphi&=&\mp\2e^{\mp i(k+1)\varphi}\big(\sin\smallover{\theta}/{2}\big)^{k+1}\big(\cos\smallover{\theta}/{2}\big)^{
2|q|-1-k}E_{\mp\alpha},
\end{array}
\label{6.9}
\eeq
where $0\leq k\leq2|q|-2$ and the upper (respectively lower) sign refers to the $a_\bz$ and $a_z$.

The geometric meaning \cite{5} of the expression (\ref{6.8}) is that they are holomorphic and antiholomorphic sections (also called ``monopole harmonics'' \cite{27}) of line bundles over the two-sphere with Chern class $2|q|-2$ [the $(-2)$ comes from the fact that our variations are vectors rather than just functions]. This is not a coincidence, since these
holomorphic sections  of line bundles are exactly the representations of the rotation group $\SU(2)$.

Now we turn to the remaining eigenspace of the Hessian. From (\ref{5.7}) and (\ref{5.10}) one sees that they are just the eigenspaces of $\bJ^2$ modulo zero modes.
Hence it suffices to consider the eigenspace of $\bJ^2$ i.e; the weights of the various representations of $\bJ^2$. Furthermore, since any weight can be obtained from the highest (or lowest) weight in a given irreducible representation $\bJ^2=j(j+1)$ by the repeated application of
$J_\pm$, it suffices to consider the highest and lower weights. However, for positive modes it turns out that the highest and lowest weights cannot be eigenvectors of
$\bb_{z\bz}$ and at the same time satisfy the divergence (zero-mode) condition $\bD\cdot\ba=0$, and since we should like to have $\bb_{z\bz}$
diagonal because it occurs (linearly) in the Hessian, we are forced at this point to drop the divergence condition. For definiteness let us therefore consider a variation of the form $(a_\bz,a_z)=(fE_\alpha,0)$ in (\ref{6.4a}),
and require it be a highest weight.

On writing the stereographic coordinate $z$ in terms of cartesian coordinates one finds that $z=x_+(r+x_3)^{-1}$ [where $x_\pm=x_1\pm ix_2$] and from this expression we see that the cartesian components of the variation $\ba$ are
\beq
(a_1,a_0,a_{-1})=
(\p z/\p x_+,\p z/\p x_3,\p z/\p x_-)a_{\bz}=
\2(f,-2zf,-z^2f)E_\alpha.
\label{6.10}
\eeq

On the other hand, one can compute from (\ref{6.2}) the Cartesian components
of the angular momentum $M$ in stereographic coordinates, and one finds that
\beq\begin{array}{c}
M_3=z\p-\bz\bp+Q,
\\[6pt]
M_+=\bp+z^2\p+Qz=\varrho\,\bp+zM_3,
\\[6pt]
M_-=-(\p+{\bz}^2\bp-Q\bz)=-\varrho\,\p+\bz M_3
\end{array}\label{6.11}
\eeq
(A simple check on (\ref{6.11}) is to note that it satisfies the usual $\so(3)$ commutation relations, that $M_\pm$ are conjugate in the Cartesian
measure $\varrho^{-2}dzd\bz$, and that $\bx\cdot\bM/r=Q$).
In terms of the Cartesian vectors (\ref{6.10}) and (\ref{6.11}) the highest weight conditions are evidently
\beqa
J_3\ba_\lambda=j\ba_\lambda
&\Rightarrow&
M_3\ba_\lambda=(j-\lambda)\ba_\lambda,
\qquad
\lambda=(-1,0,1),
\label{6.12a}
\\[6pt]
J_+\ba_\lambda=0
&\Rightarrow&
M_+\ba_\lambda=\epsilon_{\lambda}\ba_{\lambda+1},
\qquad
\epsilon_{\lambda}=(0,2,-1),
\label{6.12b}
\eeqa
where $\epsilon_{\lambda}$ are the matrix elements of $S_+=S_1+iS_2$ (and take the value shown,
rather than the conventional $\sqrt{2}\,(0,1,-1)$, because of the relative
normalization of the $\ba_\lambda$).
On inserting (\ref{6.10}) in (\ref{6.12a}) one sees that these three
equations collapse to the single equation
\beq
M_3f=(z\p-\bz\bp-|q|)f=(j-1)f
\quad\Rightarrow\quad
f={\bz}^{j+|q|-1}\Phi(\varrho),
\label{6.13}
\eeq
and on inserting this result in (\ref{6.12b}) one sees that 
$\varrho\,\bp\to z\varrho\p_{\varrho}$, and that the latter
three equations collapse to the single equation
\beq
(\varrho\p_{\varrho}+M_3+2)\Phi(\varrho)=
(\varrho\p_{\varrho}+j+1)\Phi(\varrho)=0
\quad\Rightarrow\quad
\Phi(\varrho)=\varrho^{-(j+1)}.
\label{6.14}
\eeq

Thus finally the highest weight state is $(a_\bz,a_z)=
({\bz}^{j+|q|-1}\varrho^{-(j+1)}E_\alpha,0)$. The corresponding lowest weight $(a_\bz,0)$ and the highest and lowest weights
for $(0,a_z)$ can then be read off from $(a_\bz,0)$ by using the symmetry transformations
$M_3\to-M_3$, $M_+\to-M_-$ and $q\to-q$ respectively, and thus, finally, (for $q<0$) the lowest to highest weights
for a given $j$ are
\beq
\frac{1}{(1+z\bz)^{j+1}}\left\{
\barr{c}
{\bz}^{j-|q|+1}E_\alpha\\[6pt] 0
\earr
\dots
\barr{c}
{\bz}^{j+|q|-1}E_\alpha\\[6pt] 0
\earr,
\barr{c}
0\\[6pt]
{z}^{j+|q|-1}E_{-\alpha}
\earr
\dots
\barr{c}
0\\[6pt]
{z}^{j-|q|+1}E_{-\alpha}
\earr
\right\}.
\label{6.15}
\eeq
Notice that the lowest weights in (\ref{6.15}) are \emph{not} the complex conjugates of the highest weights in the same irreducible $\bJ$ representation, but that the two representations
are conjugate. One sees that the lowest and highest weight zero modes of (\ref{6.8}) are recovered for $j=q-1$ and that these are the only modes that satisfy the divergence condition $d_za_\bz+d_\bz a_z=0$. Thus all other modes are mixtures of physical and gauge (zero-mode) states.

We should like to conclude this section by showing that the instability index $2|q|-1$ is also the Witten index for supersymmetry and the Atiyah-Singer index for the Dirac operator. Indeed, let us consider the part of $\delta^2E_\Omega$ of the Hessian, which played a central role in Sec. 5. From Eq. (5.9) one may write
$\delta^2E_\Omega=\displaystyle\int r^2drK$, where, in stereographic coordinates,
\beqa
K&=&\int dxdy g^{1/2}\tr\big\{(\varepsilon^{\alpha\beta}
d_\alpha a_\beta)^2+(g^{-1/2}d_\alpha\sqrt{g}\,g^{\alpha\beta}a_\beta)^2\big\}\nn
\\[8pt]
&=&\int dxdy g^{-1/2}\tr\big\{(d_xa_y-d_ya_x)^2
+(d_xa_x+d_ya_y)^2\big\}\nn
\\[8pt]
&=&\int dxdy\varrho^{2}\tr\big|d_za_\bz\big|^2
=\int dxdy\varrho^{2}\tr\big|d_\bz a_z\big|^2.
\label{6.16}
\eeqa
Taking half the sum of the two complex expressions in 
(\ref{6.16}), we get a \emph{supersymmetric} expression
\beq
 K=\int dxdy\varrho^2\tr(\Psi,\overline{H\Psi}),
\qquad\hbox{with Hamiltonian}\qquad
H=-\2\big\{Q^+,Q^-\big\},
\eeq
where
\beq
Q^+=\barr{cc}0&d_\bz\\ 0&0\earr,
\quad
Q^-=\barr{cc}0&0\\ d_z&0\earr,
\quad
\Psi=\barr{c}a_\bz\\ a_z\earr.
\label{6.17}
\eeq
The multiplicity $\nu$ of the ground state, which is the square integrable solution of
\beq
Q^+\Psi=\barr{c}d_\bz a_z\\ 0\earr=0,
\qquad
Q^-\Psi=\barr{c}0\\ d_z a_\bz\earr=0,
\label{6.18}
\eeq
is called the \emph{Witten index}. But these are just the negative-mode equations (\ref{6.5}). The result 
$\nu=2|q|-1$ is consistent with that found in Ref. \cite{13}.

Observe that the supersymmetric Hamiltonian $H$ can also be
written as
\beq
H=-\2{\Dir}^2,
\quad\hbox{where}\quad
\Dir=d_\bz\sigma_++d_z\sigma_-=
\barr{cc}0&d_\bz\\ d_z&0\earr
\label{6.19}
\eeq
is a Dirac-type operator, and the negative modes are exactly those satisfying
\beq
\Dir\Psi=0.
\label{6.20}
\eeq 

The number of solutions is the Atiyah-Singer (AS) index. Note, however,
that since $\ba$ is supposed to be a $2$-vector, the instability index is the $AS$ index for \textit{vectors}.
The result  $2|q|-1$ is obtained by the same calculation as the one in Atiyah and Bott \cite{6}. The advantage of this latter approach is that it generalizes to an arbitrary Riemann surface.

%%%%%%%%%%%%%%%%%%%%%%%%%%%%%%%%%%%%%%%%%%%%%%%%%
\section{LOOPS}
%%%%%%%%%%%%%%%%%%%%%%%%%%%%%%%%%%%%%%%%%%%%%%%%%
Let us consider $\Omega=\Omega(H)$, the space of loops in a compact Lie group $H$, which start and end at the identity element of $H$. The \textit{energy} of a loop $\gamma(t)$ is given by
\beq
L(\gamma)=\frac{1}{4\pi}\int_0^1\tr\big(\gamma^{-1}
\frac{d\gamma}{dt}\big)^2dt.
\label{7.1}
\eeq
A variation of $\gamma(t)$ is a 2-parameter map
$\alpha(s,t)$ into $H$ such that $\alpha(0,t)=\gamma(t)$. We fix the end points, $\alpha(s,0)=\gamma(0)$ and 
$\alpha(s,1)=\gamma(1)$ for all $s$. For each fixed $t$,
$\p\alpha/\p s$ at $s=0$ is then a vector field $X(t)$ along $\gamma(t)$, $X(0)=X(1)=0$.
$\Omega$ can be viewed then as an infinite dimensional manifold whose tangent space at a 
``point'' $\gamma$ (i.e., a loop $\gamma(t),\ 0\leq t\leq1$)
is a vector field $X(t)$ along $\gamma(t)$, which vanishes
at the end points. Since the Lie algebra $\gh$ of $H$ can
be identified with the left-invariant vector fields on $H$,
it is convenient to consider $\eta(t)=\gamma^{-1}(t)X(t)$, which is a loop in the Lie algebra,
$\gh$, s.t. $\eta(0)=\eta(1)=0$.
This is true in particular for $\zeta(t)=\gamma^{-1}(t)\frac{d\gamma}{dt}$.

The first variation of the loop-energy functional (\ref{7.1}) is
\beq
\delta L(\eta)=-\frac{1}{2\pi}\int\tr\left\{(\frac{d\zeta}{dt})\eta(t)\right\}dt.
\label{7.2}
\eeq
The \textit{critical points} of the energy satisfy therefore 
${d\zeta}/{dt}=0$, and are hence,
\beq
h(t)=\exp\big[4\pi iQt\big],
\qquad
0\leq t\leq 1,\quad
Q\in\gh
\label{7.3}
\eeq
i.e. \textit{closed geodesics} in $H$ which start at the
identity element. In order to define a loops, $Q$ must be quantized, $\exp4\pi i Q=1$. The energy of such a geodesic is obviously $L(h)=4\pi\tr(Q^2)$.

The stability properties are determined by the Hessian. After
partial integration, this is found to be
\beq
\2\delta^2L(\eta,\eta)=-\frac{1}{4\pi}\int\tr\left\{
\big(\frac{d^2\eta}{dt^2}+4\pi i[Q,\frac{d\eta}{dt}]\big)\eta\right\}dt.
\label{7.4}
\eeq
The spectrum of the Hessian is obtained hence by solving
\beq
\frac{d^2\eta}{dt^2}+4\pi i[Q,\frac{d\eta}{dt}]=\lambda\eta,
\qquad
\eta(0)=\eta(1)=0.
\label{7.5}
\eeq
Taking $\eta$ parallel to the step operators $E_{\pm\alpha}$
(\ref{7.5}) reduces to the scalar equations
\beq
\frac{d^2\eta}{dt^2}\pm4\pi iq\frac{d\eta}{dt}=-\lambda\eta,
\qquad
\eta(0)=\eta(1)=0,
\label{7.6}
\eeq
where $q=q_\alpha=\alpha(Q)$, and whose solutions yield
\beq
\eta_\alpha^{\ k}(t)=e^{\mp2\pi iqt}\big(
e^{i\pi(k+1)t}-e^{-i\pi(k+1)t}\big)E_{\pm\alpha},
\qquad
\lambda=-\pi^2\big(4q^2-(k+1)^2\big),
\label{7.7}
\eeq
where $k\geq0$ is an integer. (For $k=-1$, we
would get $\eta=0$, and for $(-k-2)$ we would
get $(-\eta_\alpha^{\ k})$). $\lambda$ is negative
if $0\leq k\leq2|q|-2$, providing us with $2(2|q_\alpha|-1)$ negative modes. The total number of negative modes is therefore the same as for a monopole with
non-Abelian charge $Q$ i.e., (\ref{5.1}) \cite{6,7}.

For $k+1=2|q|$ we get \emph{zero modes},
\beq
\eta(t)=\pm(1-e^{-4\pi i|q|t})E_{\pm\alpha},
\label{7.8}
\eeq
while for $|k|\geq2|q|$ (\ref{7.7}) yields positive modes.

If $|q|\leq1$ i.e. $q=0$ or $\pm1$, there are no negative
modes: the geodesic is stable. The results of Sec. 4 imply therefore that in each homotopy sector there is a unique stable geodesic.

Loops in $H$ can be related to YM on $\IS^2$ \cite{7}.
Indeed, the map (\ref{3.10}) i.e.
\beq
h^{\bA}(\varphi)={\cal P}\left(\exp\oint_{\gamma_{\varphi}}\bA\right)
\label{7.9}
\eeq
associates a loop $h^{\bA}(\varphi)$ to each $YM$ field
$\bA$ on $\IS^2$ \cite{2,3,8}.

For a generic connection the notation (\ref{7.9}) is merely symbolical. It can be calculated, however, explicitly if $\bA$ is Abelian, in particular, if it is a solution to the Yang-Mills equations on $\IS^2$, when it is just the geodesic (\ref{7.3}). We conclude that the map
(\ref{7.9}) carries the critical points of the YM functional
into critical points of the loop-energy functional, and that number of negative YM modes is
the same as the number of negative loop-modes. The energies of critical points are also the same, namely
$4\pi\tr(Q^2)$.

The differential of the map (\ref{7.9}) carries a YM variation $\ba$ into a loop-variation $\eta^{\bf A}(\varphi)$ i.e. a loop in the Lie algebra $\gh$.
Explicitly, let us consider 
\beq
g(\theta,\varphi)={\cal P}\left(\exp\displaystyle{\left\{
\int_0^\theta\right\}
}_{\gamma_{\varphi}}\bA\right).
\label{7.10}
\eeq
A YM variation $\ba$ goes then \cite{7} into
\beq
\eta^{\bA}(\varphi)=-\oint g^{-1}(\theta,\varphi)
a_{\theta}\big(\gamma_{\varphi}(\theta)\big)
g(\theta,\varphi)d\theta\, .
\label{7.11}
\eeq
Remarkably, $\eta^{\bA}(\varphi)$ depends on the choice of the loops $\gamma_\varphi(\theta)$ and even of the stating point. For example, with the choice of Sec. 3, the image of the YM negative mode $\ba^{(k)}$ is
\beq
\eta^{\bA}(t)=C^k(1-e^{-2\pi i kt}),
\label{7.12}
\eeq
where the numerical factor $C^k$ is,
\beq
C^k=\int_0^\pi\!(\sin\theta/2)^k(\cos\theta/2)^{2|q|-2-k}d\theta
=\frac{\Gamma\big(\frac{k+1}{2}\big)\Gamma\big(\frac{2|q|-k-1}{2}\big)}{
\Gamma(\frac{2|q|+1}{2})}\ .
\label{7.13}
\eeq
(\ref{7.12}) is similar to, but still different from the loop-eigenmodes (\ref{7.7}). If we choose however $\gamma_\varphi(\theta)$ to be the loop which starts from the south pole, goes to the north pole along the meridian at $\varphi/2$, and returns to the south pole along the meridian at
$-\varphi/2$, we do obtain (\ref{7.7}).

The map (\ref{7.9}) YM $\to$ $\big\{$loops$\big\}$
is \emph{not} one-to-one. One possible inverse
of it is given as \cite{7}
\beq
A_\theta=0,\qquad
A_\varphi=\left\{
\begin{array}{c}\smallover1/4
\displaystyle{(1-\cos\theta)}\,h^{-1}\frac{dh}{d\varphi}
\\[14pt]
-\smallover1/4
\displaystyle{(1+\cos\theta)}\,\frac{dh}{d\varphi}h^{-1}
\end{array}\right.
\qquad\hbox{in}\qquad
\left\{
\begin{array}{c}
N
\\[24pt]
S
\end{array}\right.
\label{7.14}
\eeq

%%%%%%%%%%%%%%%%%%%%%%%%%%%%%%%%%%%%%%%%%%%%%%%%%
\section{GLOBAL ASPECTS}
%%%%%%%%%%%%%%%%%%%%%%%%%%%%%%%%%%%%%%%%%%%%%%%%%

Into what can an unstable monopole go~? It can not leave its homotopy sector, since this would require infinite energy. But it can go into another configuration in
the same sector, because any two such configurations are separated only by finite energy. To see this one has only to note that the family of configurations
\beq
{\bA}^\tau=\tau\,\bA'+(1-\tau)\bA,
\qquad
\Phi^\tau=\Phi,
\qquad
0\leq\tau\leq1
\label{8.1}
\eeq
which are not, in general, solutions of the field equations except for $\tau=0,1$, but which interpolate smoothly between solutions $(\bA,\Phi)$ and $(\bA',\Phi)$.
They all lie in the same Higgs sector because $\Phi$ does not change, and have finite energy for all $0\leq\tau\leq1$. Indeed, as $r\to\infty$, one has
\beqa
&{\bA}^\tau\sim {1}/{r},
\qquad
&r^{3/2}(D^\tau\Phi)=\tau (r^{3/2}D'\Phi)+
(1-\tau)(r^{3/2}D\Phi)\to0,
\nn
\\[6pt]
&&r^{3}V(\Phi^\tau)=r^{3}V(\Phi)\to0,
\label{8.2}
\eeqa
so that the energy integral (\ref{2.1}) converges for
$(\bA^\tau,\Phi)$. As a matter of fact, one may obtain a rather simple and compact expression for the interpolated energy $E^\tau=E(\bA^{\tau})$ as follows:
\beq
D^\tau\Phi=\tau(D'\Phi)+(1-\tau)(D\Phi)
\quad\hbox{and}\quad
B_{ij}^{\ \tau}=\tau B'_{ij}+(1-\tau)B_{ij}+
\tau(1-\tau)[\Delta_i,\Delta_j],
\label{8.3}
\eeq
where $\Delta_j=\bA'_j-\bA_j$. This shows that the
interpolated energy must be of the general form
\beq
E^\tau=a\tau^2+b(1-\tau)^2+c\tau^2(1-\tau)^2+
2f\tau(1-\tau)+2g\tau^2(1-\tau)+2h\tau(1-\tau)^2,
\label{8.4}
\eeq
where $a,\dots, g$ are integrals over the field
configurations which are independent of $\tau$,
and in particular
\beq
a=E',\quad
b=E,\quad\hbox{and}\quad
c=\int d^3x\tr[\Delta_i,\Delta_j]^2,
\label{8.5}
\eeq
where $E$ and $E'$ are the energies of the solutions $(\bA,\Phi)$ and $(\bA',\Phi)$. But since the solutions are
extremal points of the energy, $\p E^\tau/\p\tau$ must vanish at $\tau=0,\,1$ and this leads to the conditions
$a=f+g$ and $b=f+h$. Using these two equations to eliminate $h$ and $g$ one finds that $f$ is also eliminated and thus $E^\tau$ reduces to the simple expression
\beq
E^\tau=\tau^2(3-2\tau)E'+(1-\tau)^2(1+2\tau)E
+\tau^2(1-\tau)^2c.
\label{8.6}
\eeq
Since $\Delta\sim 1/r$ as $r\to\infty$ it is evident that $c$ is finite, and hence that the
interpolated energy is finite for all $0\leq\tau\leq1$.
Thus the energy barrier between $E$ and $E'$ is finite.

Yang-Mills-Higgs theory on $\IR^3$ has the same topology as YM on $\IS^2$. The true configuration space ${\cal C}$
of this latter is furthermore the space ${\cal A}$ of all
YM potentials modulo gauge transformations,
\beq
{\cal C}\simeq{\cal A}/{\cal H}
\quad\hbox{where}\quad
{\cal H}=\big\{\hbox{Maps\;} \IS^2\to H\big\},
\label{8.7}
\eeq
and the \textit{path components} of ${\cal C}$ are just
the \textit{topological sectors}: $\pi_0({\cal C})\simeq
\pi_2(G/H)$.

When studying the topology of ${\cal C}$, we can also use loops. The map (\ref{7.9}) (widely used for describing the topological sectors \cite{2,3,8}), is, in fact a \emph{homotopy equivalence} between YM on $\IS^2$ and
$\Omega=\Omega(H)/H$, the loop-space of $H$ modulo global
gauge rotation \cite{15}. (One has to divide out by $H$ because a gauge-transformation changes the non-integrable
phase factor by a global gauge rotation). This correspondence explains also why we could use the diagram
for counting the negative YM modes, introduced by Bott \cite{12} originally for loops.

Most saddle-point solutions studied so far in field theories are associated to \textit{non-contractible loops} \cite{17}. There are no non-contractible loops in our case, $\pi_1({\cal C})\simeq \pi_1(\Omega)\simeq \pi_2(H)=0$. There are, however, \emph{noncontractible two-spheres}: $\pi_2({\cal A}/{\cal H})\simeq \pi_1({\cal H})\simeq\pi_3(H)$.
But for any compact $H$, $\pi_3(H)$ is the direct sum of the
$\pi_3$'s of the simple factors $K_j,\, j=1,\dots,s$. On the other hand, for any compact, simple Lie group, $\pi_3\simeq\IZ$, the
only exception being $\SO(4)$, whose $\pi_3$ is 
$\IZ\oplus\IZ$.

Below we associate an energy-reducing two-sphere which interpolates between a given (unstable) monopole and some other, lower energy monopole to each
intersection of the line $0\leftrightarrow Q$ with the
root plane. The tangent vectors to these spheres are furthermore negative modes for the Hessian.

The role of our spheres is explained by Morse theory \cite{16}: a critical point of index $\nu$ of a ``perfect Morse function'' is in fact associated to a class in $H_\nu$, the
$\nu$-dimensional homology group. Intuitively 
(Fig. \ref{Fig2}), following the $\nu$ independent negative-mode directions, we get a small $\nu$-dimensional ``cap'' which, when glued to the lower-energy part of configuration space, forms a closed, $\nu$-dimensional surface. The Hurewicz isomorphism \cite{25} tells however that, for simply connected manifolds, $\pi_2$ is isomorphic to $H_2$, the second homology group. The K\"unneth formula \cite{25} shows furthermore that the direct product of the $(\nu/2)$ $2$-spheres has a non-trivial class in $H_\nu$.

Let us first consider a geodesic $\exp4\pi iQt,\,
0\leq t\leq1$, rather than a monopole. Remember that the step operators $E_{\pm\alpha}$ and $H_\alpha=[E_{\alpha},E_{-\alpha}]$ close to an $\ort(3)$ subalgebra of $\gk\subset\gh$. Denote by $G_\alpha$ the generated subgroup of $K\subset H$. Our two-spheres are associated to these $G_\alpha$'s.

Observe first that, for each root $\alpha$, 
\beq
S_\alpha=\big\{g^{-1}Q_\alpha g,\,g\in G_\alpha\big\},
\eeq
is a two-sphere in the Lie algebra $\gk\subset\gh$. If 
$\xi$ is an arbitrary vector from $S_\alpha$,
\beq
\exp\big(\pi i\xi\big)=\exp(\pi i\,g^{-1}Q_\alpha g)=
g^{-1}\big(\exp[\pi iQ_\alpha]\big) g=
\pm 1,
\label{8.8}
\eeq
(the sign depends on $G_\alpha$ being $\SU(2)$
or $\SO(3)$), because $\exp 2\pi iQ_\alpha=1$.
Hence, for each $\xi$ from $S_\alpha$ and integer $k$,
\beq
h_\xi^{\ k}(t)=e^{\pi it(k+1)\xi}e^{2\pi it(2Q-(k+1)Q_\alpha/2)}
\label{8.9} 
\eeq
is a loop in $H$. Equation (\ref{8.9}) is therefore a \emph{two-sphere of loops} in $H$, parametrized
by $\xi\in\IS^2$. Using the shorthand $h=h_\xi^{\ k}$
and $\zeta=(2Q-(k+1)Q_\alpha/2)$, the speed of the loop (\ref{8.9}) is
\beq
h^{-1}\frac{dh}{dt}=e^{-2\pi i\zeta t}\big((k+1)\pi\xi\big)
e^{2\pi i\zeta t}+2\pi\zeta.
\label{8.10}
\eeq
To calculate its energy, observe that, for any vector $\zeta$ from the Cartan algebra, $\zeta-\alpha(\zeta)H_{\alpha}/
(\alpha,\alpha)$ commutes with $E_{\pm\alpha}$, because
$$
[\zeta-\alpha(\zeta)\frac{H_{\alpha}}{
(\alpha,\alpha)},E_\alpha]=
\alpha\big(\zeta-\alpha(\zeta)\frac{H_{\alpha}}{
(\alpha,\alpha)}\big)E_\alpha=0,
$$
and so
$$
g\,\zeta\,g^{-1}=g\big(\zeta-\alpha(\zeta)\frac{H_{\alpha}}{
(\alpha,\alpha)}+\alpha(\zeta)\frac{H_{\alpha}}{
(\alpha,\alpha)}\big)g^{-1}
=\zeta-\alpha(\zeta)\frac{H_{\alpha}}{
(\alpha,\alpha)}+
\big(\frac{\alpha(\zeta)}{(\alpha,\alpha)}\big)\,gH_\alpha\,g^{-1}.
$$
Hence
\begin{eqnarray*}
\tr(\xi,\zeta)&=&\tr\big(g^{-1}Q_\alpha\,g,\zeta\big)=
\tr\big(Q_\alpha,g\zeta\,g^{-1})
\\[6pt]
&=&\tr\big(Q_\alpha,\zeta-\alpha(\zeta)\frac{H_{\alpha}}{
(\alpha,\alpha)}\big)+
\frac{\alpha(\zeta)}{(\alpha,\alpha)}\tr(Q_\alpha,
g\,H_\alpha\,g^{-1}).
\end{eqnarray*}
Substituting here $\zeta$ we get finally, using $\tr(Q_\alpha Q)=\alpha(Q)=q$,
\beq
L(h)=\pi\big\{2(k+1)(2|q|-k-1)\tr(Q_\alpha/2)^2
\cos\tau+\tr(2Q-(k+1)Q_\alpha/2)^2\big\},
\label{8.11}
\eeq
where $\tau$ is the angle between $Q_\alpha$ and $g^{-1}Q_\alpha\,g$.

For $\tau=0,\,\pi$ i.e. for $\xi=\pm Q_\alpha$ the two factors in (\ref{8.10}) commute.
For $\tau=0$ we get the geodesic $\exp(4\pi iQt)$ and for
$\tau=\pi$ i.e. $\xi=-Q_\alpha$ we get another, lower-energy geodesic, namely
\beq
h_{\alpha}^{\ k}(t)=e^{2\pi it(2Q-(k+1)Q_\alpha)},
\qquad
0\leq t\leq 1.
\label{8.12}
\eeq

We conclude that, for each $0\leq k\leq2|q|-2$, (\ref{8.9}) provides us with a
smooth \textit{energy reducing two sphere of loops}, whose
top is the ``long'' geodesic we started with, and whose bottom is (\ref{8.12}).

Carrying out this construction for all roots $\alpha$
and all integers $k$ in the range $0\leq k\leq2|q|-2$, we get exactly the required number of two-spheres.
They can also be shown to be non-contractible, and to generate $\pi_2$.

Consider now the tangent vectors to our two-spheres of loops along the curves 
$$
g_s(t)=e^{2\pi i E_{\pm\alpha}s}Q_\alpha
e^{-2\pi i E_{\pm\alpha}s}
$$
at $s=0$, the top of the spheres. They are
\beq
e^{4\pi i|q|t}\big(
e^{-2\pi i(2|q|-k-1)t}-e^{4\pi i|q|t}\big)E_{\pm\alpha}.
\label{8.13}
\eeq

The loop-variations (\ref{8.13}) are again negative modes. They are not, however, eigenmodes, but rather mixtures of negative modes
$(1-e^{-2\pi i(2|q|-k-1)t})E_{\pm\alpha}$ and the
zero mode $(1-e^{-4\pi i|q|t})E_{\pm\alpha}$.

The inverse formula (\ref{7.14}) translates finally the whole construction to YM: $A_\theta^{\ \xi}=0$,
\beq
A_\varphi^{\ \xi}=\left\{\begin{array}{c}
\smallover1/4
\displaystyle{(1-\cos\theta)}\big(e^{-2\pi i\zeta t}(k+1)\xi\,e^{2\pi i\zeta t}+2\zeta\big)
\\[16pt]
-\smallover1/4
\displaystyle{(1+\cos\theta)}\big((k+1)\xi+
e^{\pi i(k+1)t\xi}(2\zeta)\,e^{-\pi i(k+1)t\xi}\big)
\end{array}\right.
\quad\hbox{in}\quad
\left\{\begin{array}{c}
N\\[22pt]
S
\end{array}\right.
\label{8.14}
\eeq
in fact a ``round'' energy-reducing two-sphere of YM
potentials on $\IS^2$. The top of the sphere is $QA_D$, the monopole
we started with, and the bottom is another, lower-energy
monopole, whose charge is $Q-(k+1)Q_\alpha/2$. Again, the situation is well illustrated on the Bott diagram, (see 
the next Section).

Note finally that our definition (\ref{8.9}) can easily be
modified so that the spheres fit the negative eigenmodes (\ref{7.7}). However, the loops are then no longer of constant speed and do not interpolate in a monotonically energy-reducing manner between the critical points.

%%%%%%%%%%%%%%%%%%%%%%%%%%%%%%%%%%%%%%%%%%%%%%%%%
\section{EXAMPLES}
%%%%%%%%%%%%%%%%%%%%%%%%%%%%%%%%%%%%%%%%%%%%%%%%%

\kikezd{Example 1}

\begin{figure}
\begin{center}\hspace{-4mm}
\includegraphics[scale=.85]{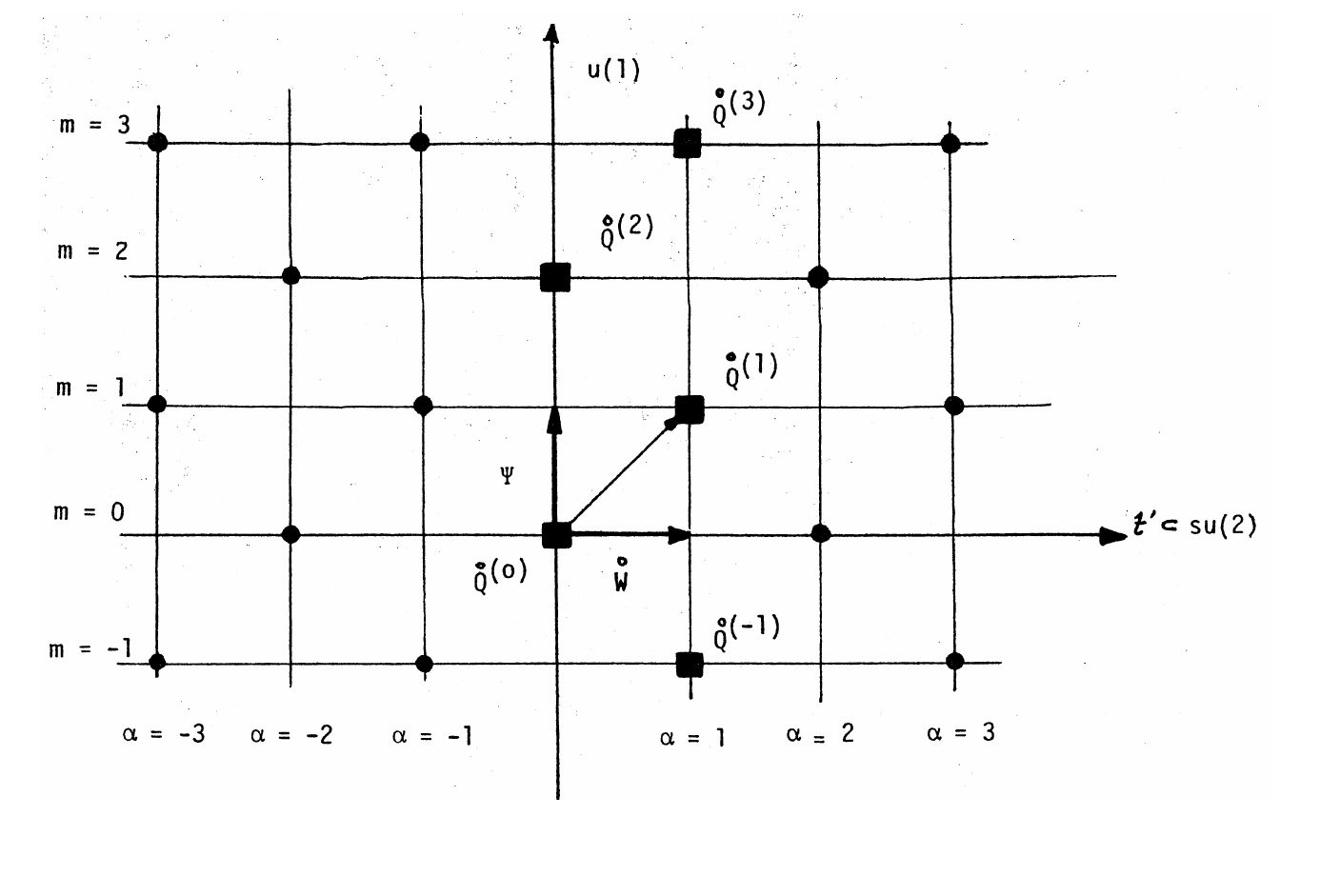}\vspace{-12mm}
\end{center}
\caption{{\it Diagram of $\UN(2)$. The horizontal axis represents the Cartan algebra of $\su(2)$, the vertical axis is the center generated by $\Psi$. The charges are $Q=m\Psi+\oW_{[m]}+n\sigma_3$, where $m$ is the
topological quantum number, $[m]=m$ (mod $2$).
$W_1=\sigma_3/2$ is a minimal vector and $Q_1=\sigma_3$ is a primitive charge. The root ``planes'' are vertical lines which intersect the horizontal axis at integer multiples of 
$\oW_1$. The horizontal lines are the topological sectors, labelled by $m$. The pattern is periodic in $m$ (mod $2$).
In each sector, the stable charge $Q$ is the one closest to the center. For example, the monopole with charge $Q=\2\sigma_3$} \cite{18} {\it lies in the vacuum sector and has $2$ independent instabilities. $Q$ is at the top of an energy-reducing $2$-sphere whose
bottom is the vacuum.}}
\label{Figure3}
\end{figure}

The simplest case of interest is that of when the little group $H$ of the Higgs field is $H=U(2)$
(Fig. \ref{Figure3}). The Cartan algebra consists of diagonal matrices (combinations of $\sigma_3$ and the unit matrix $\II_2$); the only positive root $\alpha$ is the difference of the diagonal entries. In fact,
\beq\begin{array}{lll}
E_+=\sigma_+=\barr{cc}0&1\\ 0&0\earr,
&E_-=\sigma_-=\barr{cc}0&0\\ 1&0\earr,
&H=Q=\sigma_3=\barr{cc}1&0\\ 0&-1\earr,
\\[16pt]
X=\sigma_1=\barr{cc}0&1\\ 1&0\earr,
&Y=\sigma_2=\barr{cc}0&-i\\ i&0\earr,
&
\oW_1=\2\sigma_3=\2\barr{cc}1&0\\ 0&-1\earr.
\end{array}
\label{9.1}
\eeq
The only primitive vector, $\oW_1$, is also a minimal one: in fact,
$\exp 2\pi i\oW_1=-1$. $Q_1=2\oW_1=\sigma_3$ generates the charge lattice of $K=\SU(2)$ which is also the topological zero-sector
of $\UN(2)$. The topological sectors are labelled by a single integer $m$, defined by $2Q_{||}=m{\rm diag}(1/2,-1/2)=m\Psi$. 
%%%%%%%%%%%%%%%%%%%%%%%
%%%%%%%%%%%%%%%%%%%%%%%%%%%%%%%%%%%%%%%%%%%%%%%
%%%%%%%%%%%%%%%%%%%%%%%%%%%
The unique stable monopole of the sector
$m$ is
\beq
\oQ^{(m)}=\2m\Psi+\2\oW_{[m]}=\left\{
\begin{array}{lll}
\2{\rm diag\ } (k,k)
&\hbox{for} &m=2k
\\[8pt]
\2{\rm diag\ } (k+1,k)
&\hbox{for} &m=2k+1
\end{array}\right.
\label{9.2}
\eeq
where $[m]$ is $m$ modulo $2$ and $\oW_\theta=0$ by convention.
Any other monopole of the sector $m$ is of the form
\beq
Q^{(m)}=\oQ^{(m)}+\frac{n}{2}Q_1=\oQ^{(m)}+\frac{n}{2}\sigma_3
=\oQ^{(m)}+\2{\rm diag }(n,-n).
\label{9.3}
\eeq
Those monopoles for which $n\neq0$ are unstable, with index $\nu=2(2n-1)$ for $m$ even and $\nu=4n$ for $m$ odd. For example,
when $G=\SU(3)$ is broken to $H=\UN(2)$ by an adjoint Higgs $\Phi$, the vacuum sector contains a configuration whose non-Abelian
charge $Q$ is conjugate to ${\rm diag\ }(1/2,-1/2,0)$ \cite{18}.
Our result shows that this configuration is, (as conjectured), unstable, and has rather $2$ negative modes, namely
\beq
a_\theta=\2e^{\mp i\varphi}\,\sigma_{\pm},
\qquad
a_\varphi=\mp\2e^{\mp i\varphi}\sin\theta\,\sigma_{\pm}.
\label{9.4}
\eeq

The construction of Sec. 8 yields, furthermore, an energy reducing $2$-sphere of YM configuration, namely 
$A_\theta^{\ \xi}=0$,
\beq
A_\varphi^{\ \xi}=\left\{\begin{array}{c}
\smallover1/4
\displaystyle{(1-\cos\theta)}\big(e^{-i\varphi\2\sigma_3}
\,\xi\,e^{i\varphi\2\sigma_3}+\sigma_3\big)
\\[12pt]
-\smallover1/4
\displaystyle{(1+\cos\theta)}\big(\xi+
e^{i\varphi\2\xi/2}\,\sigma_3\,e^{-i\varphi\xi/2}\big)
\end{array}\right.
\quad\hbox{in}\quad
\left\{\begin{array}{c}
N\\[18pt]
S
\end{array}\right.
\label{9.5}
\eeq
where $\xi=g^{-1}\sigma_3\,g$. Parametrizing this two-sphere with Euler angles $(\tau,\varrho)$ (say), we can write
\beqa
\big(e^{-i\varphi\sigma_3}\big)\xi\big(e^{i\varphi\sigma_3}\big)
&=&
\cos\tau\sigma_3-\sin\tau\big\{e^{-i(\varrho+\varphi)}\sigma_+
+e^{i(\varrho+\varphi)}\sigma_-\big\}=\nn
\\[8pt]
&=&\cos\tau\sigma_3-\sin\tau\big\{\cos(\varrho+\varphi)\sigma_1+\sin(\varrho+\varphi)\sigma_2\big\},
\label{9.6}
\eeqa
so that the $2$-sphere (\ref{9.5}) becomes
\beqa
A_\theta^{\ \xi}=0,\qquad
A_\varphi^{\ \xi}=\smallover1/4(1-\cos\theta)\left\{
(1+\cos\tau)\sigma_3-\sin\tau(e^{-i(\varrho+\varphi)}\sigma_++
e^{i(\varrho+\varphi)}\sigma_-)\right\}\nn
\\[8pt]
=\smallover1/4(1-\cos\theta)\left\{
(1+\cos\tau)\sigma_3-\sin\tau\big(\cos(\varrho+\varphi)\sigma_1
+\sin(\varrho+\varphi)\sigma_2\big)\right\}
\label{9.7}
\eeqa
in $N$, and similarly in $S$.  For $\tau=0,\ \xi=\sigma_3$ we get $\bA=(\sigma_3/2)\bA^D$; i.e. the monopole
we started with and for $\tau=\pi,\ \xi=-\sigma_3$ we get $\bA=0$,
the vacuum. The energy of the configuration (\ref{9.5}) is
\beq
E^{(\tau,\varrho)}=\pi(1+\cos\tau),
\label{9.8}
\eeq
which is consistent with (\ref{8.11}). Observe that 
$(1+\cos\tau)$ is just the height function on the unit sphere.

%%%%%%%%%%%%%%%%%%%%
\kikezd{Example 2}
%%%%%%%%%%%%%%%%%%%

The physically most relevant example is when the Higgs little
group is $H=\UN(3)$ i.e. locally $\su(3)_c+\un(1)_{em}$, the symmetry group of the strong and electromagnetic interaction.

\begin{figure}
\begin{center}\hspace{-4mm}
\includegraphics[scale=.9]{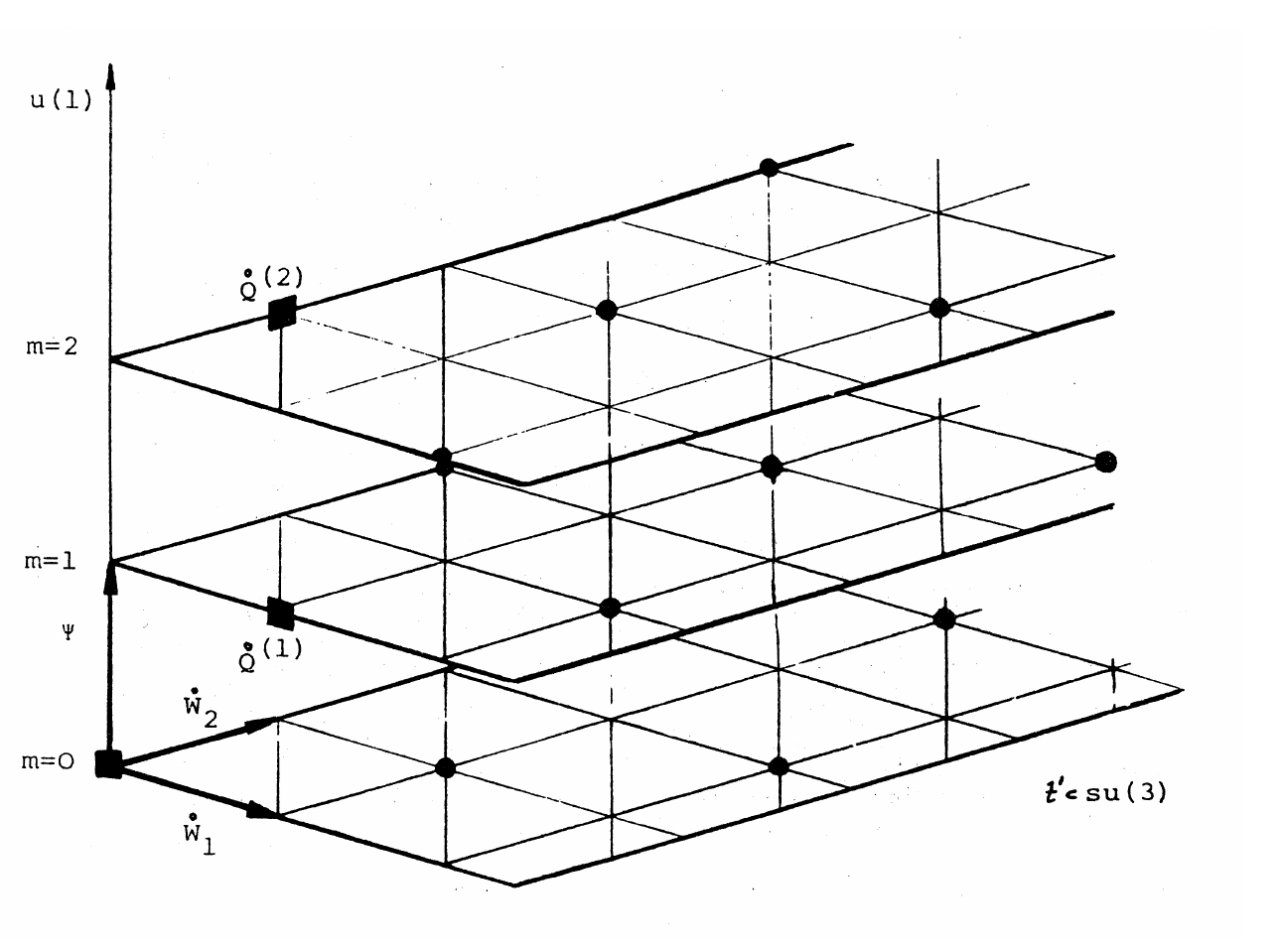}\vspace{-9mm}
\end{center}
\caption{\it Diagram of $\UN(3)$. The vertical axis represents the center, generated by $\Psi={\rm diag }(1/3,1/3,1/3)$, and the horizontal plane $\gt'$ is the Cartan algebra of $\SU(3)$, shown in more detail on Fig. \ref{Figure5}. The charges are
$m\,\Psi+\oW_{[m]}+n_1Q_1+n_2Q_2$, where $Q_1$ and $Q_2$ are
the primitive charges of $\SU(3)$. 
 The horizontal planes are the topological sectors. Sector $m$ is obtained from the vacuum sector by shifting by $m\,\Psi+\oW_{[m]}$. In each sector, the charge closest to the center is that of the unique stable monopole.
The diagram is periodic in $m$ modulo $3$.}
\label{Figure4}
\end{figure}

The diagram is now three-dimensional, the central $\un(1)$ being
the vertical axis on Fig. \ref{Figure4}
%%%%%%%%%%%%%%%%%%%%%%
%%%%%%%%%%%%%%%%%%%
and $\gt'$ being the horizontal plane. The primitive roots are $\alpha_1(X)=X_1-X_2$ and $\alpha_2(X)=X_2-X_3$ 
(for $X={\rm diag}(X_1,X_2,X_3)$). The corresponding primitive vectors, 
$\oW_1={\rm diag\ }(2/3,-1/3,-1/3)$ and 
$\oW_2={\rm diag\ }(1/3,1/3,-2/3)$, are also minimal vectors:
their exponentials are in bijection with the elements in the
$\IZ_3$-center of $\SU(3)$.

The highest root is $\theta=\alpha_1+\alpha_2$, and the charge lattice
of $K=\SU(3)$ is generated by $Q_1={\rm diag}(1,-1,0)$ and
$Q_2={\rm diag}(0,1,-1)$.

The topological sectors are labelled by an integer $m$.
In fact, the projection of Sector $m$ onto the center is
$m\,\Psi=m\,{\rm diag}(1/3,1/3,1/3)$. The unique stable charge in
this sector is
\beq
\oQ^{(m)}=m\Psi+\oW_{[m]}=\left\{\begin{array}{l}
{\rm diag}(k,k,k)
\\[6pt]
{\rm diag}(k+1,k,k)
\\[6pt]
{\rm diag}(k+1,k+1,k)
\end{array}\right.
\qquad
\hbox{for}
\qquad
m=
\left\{\begin{array}{l}
3k
\\[6pt]
3k+1
\\[6pt]
3k+2
\end{array}\right.
\label{9.9}
\eeq
where $[m]$ means $m$ modulo $3$. Any other monopole has charge
\beq
Q=\oQ+Q'=
\oQ+n_1\frac{Q_1}{2}+n_2\frac{Q_2}{2}=
\oQ+\2{\rm diag}(n_1,n_2-n_1,-n_2).
\label{9.10}
\eeq

Those configuration with $Q'\neq0$ are unstable.

\begin{figure}
\begin{center}\hspace{-4mm}
\includegraphics[scale=0.8]{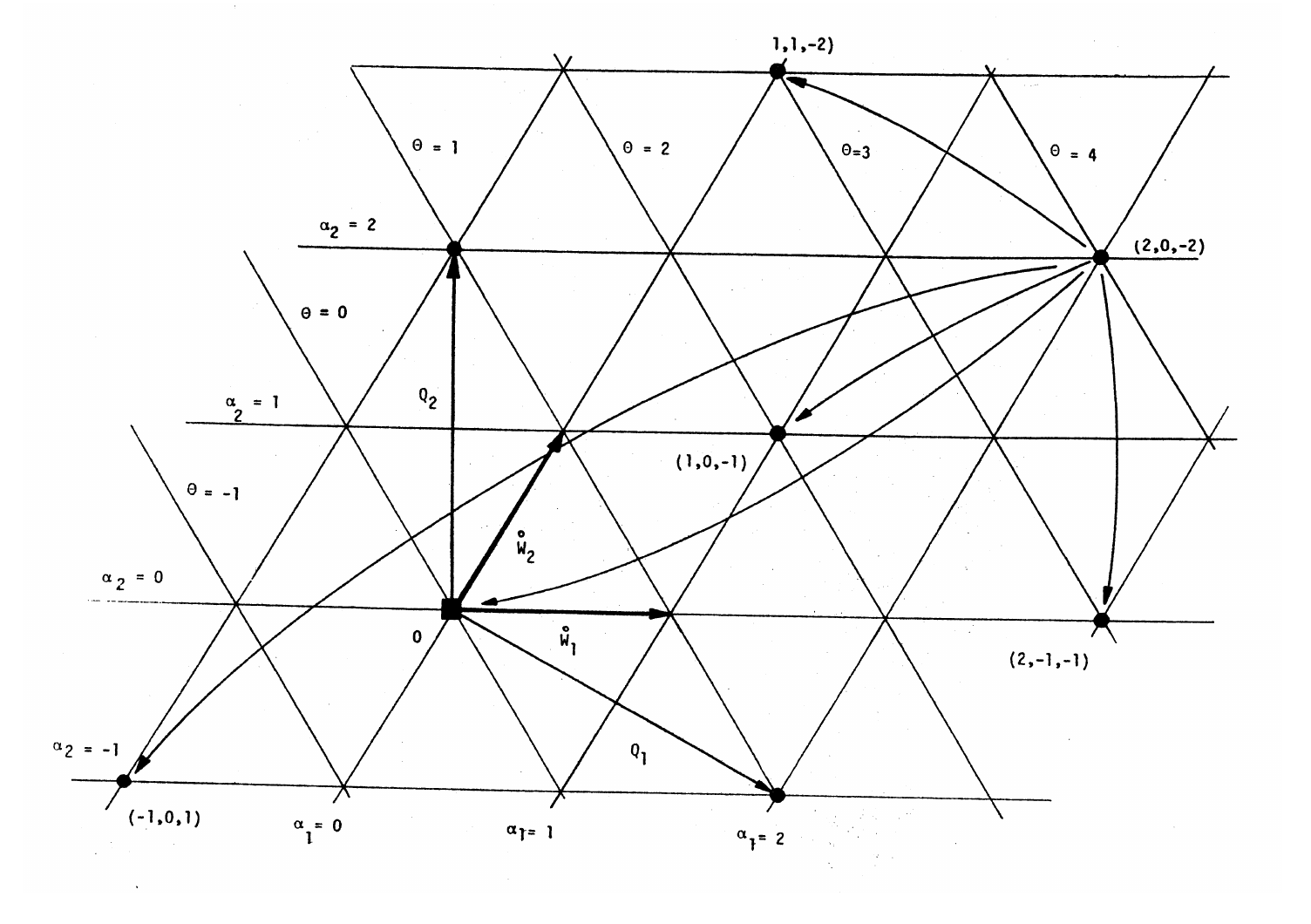}\vspace{-9mm}
\end{center}
\caption{\it Diagram of $\SU(3)$. $Q_1={\rm diag}(1,-1,0)$ and $Q_2={\rm diag}(0,1,-1)$ are the primitive charges and the two primitive roots are $\tr(Q_1\,\cdot\,)$
and $\tr( Q_2\,\cdot\,)$. 
The minimal vectors 
$\oW_1=\smallover1/3{\rm diag}(2,-1,-1)$ and
$\oW_1=\smallover1/3{\rm diag}(1,1,-2)$ generate the diagram. There are three root planes, intersecting in angle $\pi/3$. For example, the monopole whose charge is 
$2Q={\rm diag}(2,0,-2)$ has $10$ negative modes, tangent to $5$ energy-reducing $2$-spheres, which end at $(2,-1,-1),\, (1,1,-2),\, (1,0,-1),\, (0,0,0),\ (-1,0,1)$.}
\label{Figure5}
\end{figure}

For example, the $Q={\rm diag}(1,0,-1)$ (Fig. \ref{Figure5})
belongs to the vacuum sector, because its charge is in $\gk=\su(3)$. 
$$
\alpha_1(2Q)=2,\quad
\alpha_2(2Q)=2,\quad
\theta(2Q)=4,
$$
and so there are $10$ negative modes, given by (\ref{6.18}).
Equation (\ref{8.3}) yields in turn $5$ energy-reducing
$2$-spheres, which end at
\beqa
Q_{\alpha_1}={\rm diag\ }(1,-1/2,-1/2),
\qquad
Q_{\alpha_2}={\rm diag\ }(1/2,1/2,-1),
\label{9.11}
\\[6pt]
Q_{\theta}^1={\rm diag\ }(1/2,0,-1/2),\quad
Q_{\theta}^2=0,\quad
Q_{\theta}^3={\rm diag\ }(-1/2,0,1/2),
\nn
\eeqa
%%%%%%%%%%%%%%%%%%%%%%
%%%%%%%%%%%%%%%%%%%%%%%%%%%
%%%%%%%%%%%%%%%%%%%%%%%%%

%%%%%%%%%%%%%%%%%%%%
\kikezd{Example 3}
%%%%%%%%%%%%%%%%%%%

In Ref. \cite{19} the authors consider a $6$-dimensional pure
$\SU(3)/\IZ_3$ Yang-Mills model, defined over $M^4\times\IS^2$,
where $M^4$ is Minkowski space. They claim that any
(Poincar\'e)$\times\SO(3)$ symmetric configuration is unstable against the formation of tachyons.

A counterexample is given by Forgacs \textit{et al}. \cite{20},
who show that the ``symmetry-breaking vacuum''
\beq
A_i=0,\quad i=1,\dots4,
\qquad
\bA=\smallover1/6\,{\rm diag}(2,-1,-1)\,\bA^D
\label{9.12}
\eeq
(where $\bA$ is a $2$-vector on the extra-dimensional $\IS^2$), is stable.

\begin{figure}
\begin{center}\vspace{-2mm}
\includegraphics[scale=0.8]{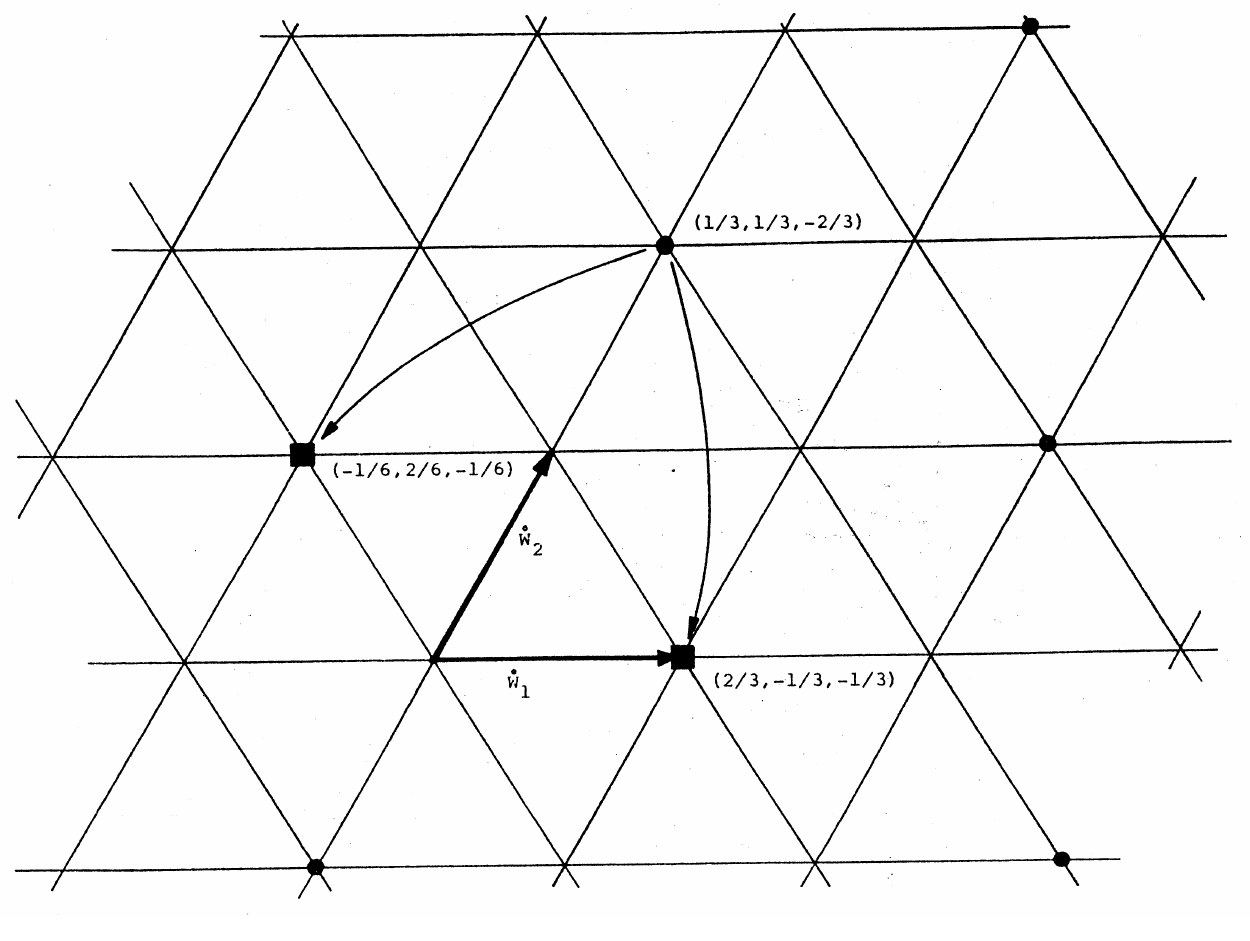}\vspace{-5mm}
\end{center}
\caption{{\it  Diagram of $\SU(3)/\IZ_3$, the adjoint group of $\SU(3)$. The diagram is essentially identical to that of $\SU(3)$, the only difference being that the primitive $W$'s are now charges. In fact,
$\oW_j,\ j=0,1,2,$ are the stable charges of the three topological sectors.} ({\it Only the sector $\exp[4\pi iQ]=
e^{4\pi i/3} $ is shown.}) {\it For example, $Q=2\oW_2={\rm diag}(1,-\2,-\2)$ is unstable with $4$ independent negative modes. It lies
at the top of two energy-reducing $2$-spheres, whose bottoms are $\oW_1$ and its conjugate.}}
\label{Figure6}
\end{figure}

These observations have a simple explanation: the assumption of spherical symmetry in the extra dimensions leads to asymptotic monopole configurations on $\IS^2$ with gauge group 
$H=\SU(3)/\IZ_3$. Since $\pi_1(\SU(3)/\IZ_3)\simeq\IZ_3$,
there are three topological classes corresponding to the three central elements
$z^*_0=1,\, z^*_1=e^{2\pi i/3},\,z^*_2=e^{4\pi i/3} $ of
$H=\SU(3)$ (Fig. \ref{Figure6}).
%%%%%%%%%%%%%%%%%%%%%%%%%%%%
%%%%%%%%%%%%%%%%%%
The diagram of $H=\SU(3)/\IZ_3$ differs from that of $H=\SU(3)$
only in that the primitive $W_i$'s are already charges in this case.

(\ref{9.12}) is indeed stable, because it is an asymptotic monopole with charge $\oQ=\2\oW_1$, the unique stable charge of the Sector characterized by $z^*_2$. On the other hand,
any other configuration, e.g. \cite{20}
\beq
\bA=\smallover1/3{\rm diag}(1,1,-2)\,\bA^D
\label{9.13}
\eeq
is unstable. Counting the intersections with the root planes shows that there are $\nu=4$ negative modes.

Both configuration (\ref{9.12}) and (\ref{9.13}) belong to the
same sector, and the construction of Sec. 8 provides us with two energy-reducing two-spheres from the
monopole (\ref{9.13}) to those
with charges $\smallover1/6{\rm diag}(2,-1,-1)$ (i.e. (\ref{9.12})) and
its conjugate $\smallover1/6{\rm diag}(-1,2,-1)$.

Choosing rather $\oQ=\2\oW_2$ in (\ref{9.12}) would obviously lead
to another stable configuration.

%%%%%%%%%%%%%%%%%%%%
\kikezd{Example 4}
%%%%%%%%%%%%%%%%%%%

\begin{figure}
\begin{center}\hspace{-4mm}
\includegraphics[scale=0.84]{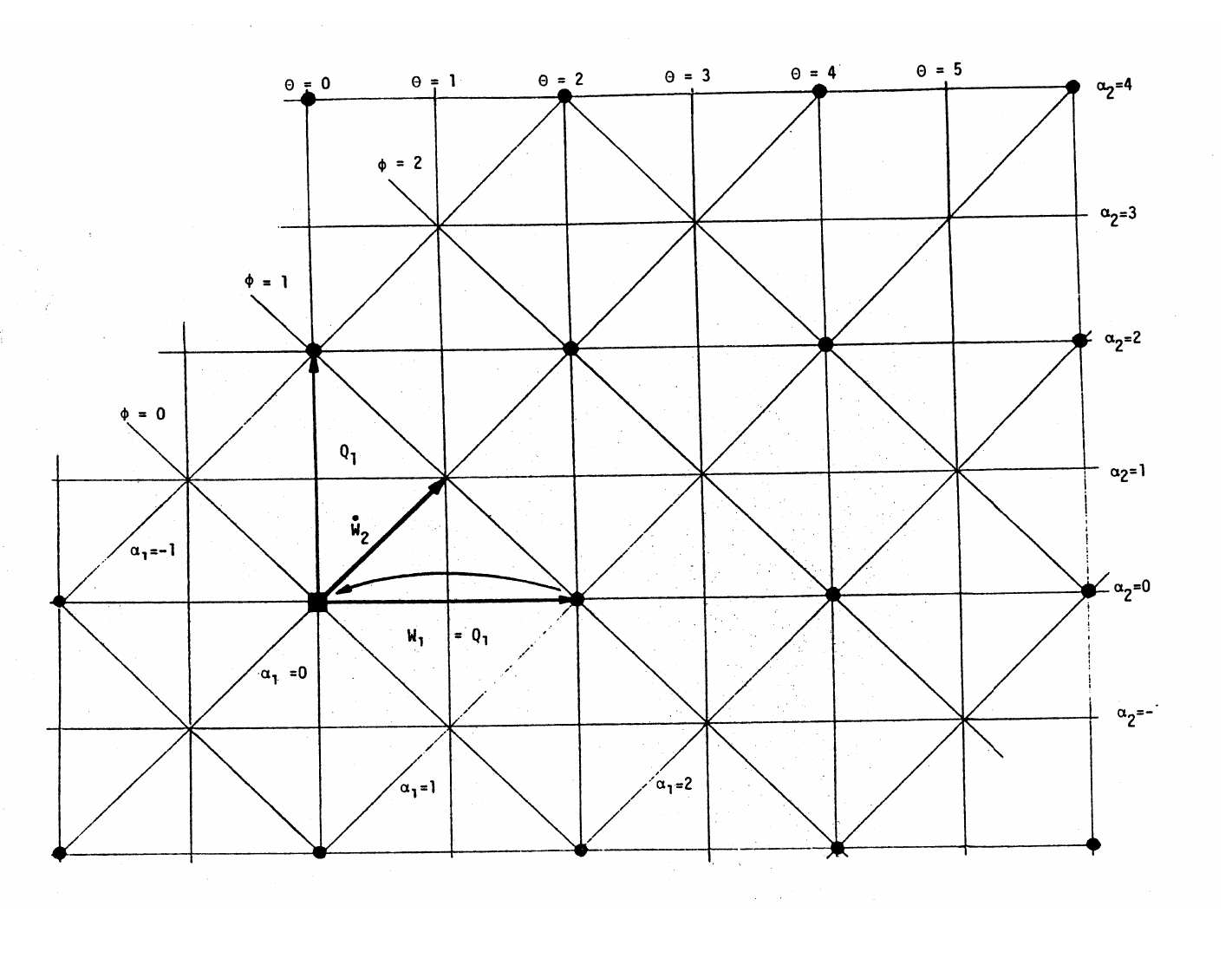}\vspace{-10mm}
\end{center}
\caption{\it Diagram of ${\rm Sp}(4)\approx{\rm Spin}(5)$, the double covering of $\SO(5)$. The primitive
charges are $Q_1={\rm diag}(1,0,-1,0)$ and $Q_2={\rm diag}(0,1,0,-1)$. The two primitive $W$'s are 
$W_1={\rm diag}(1,0,-1,0)$ and  
$\oW_2={\rm diag}(\2,\2,-\2,-\2)$ out of which only $\oW_2$ is minimal. There are $4$ families of root planes. The monopole with charge
$2Q=W_1$ is unstable with two negative modes. It lies on the top of an energy-reducing $2$-sphere which ends at the vacuum.
}
\label{Figure7}
\end{figure}

To have a simple example where not all primitive weights are minimal, let us assume that the residual group is
$$
H=\big(\UN(1)\times{\rm Sp}(4)\big)/\IZ_2.
$$
Then $\gk={\rm sp}(4)=\so(5)$ and $K$ is ${\rm Spin}(5)$, the
double covering of $\SO(5)$. $\gk$ can be represented by
$4\times4$ symplectic matrices with a $2$-dimensional Cartan algebra, say $\gt'={\rm diag}(a,b,-a,-b)$. The charge
lattice consists of those vectors in $\gt'$ with integer entries. Let us choose the primitive roots $\alpha_1=\tr(H_1\,\dot\,)$
and $\alpha_2=\tr(H_2\,\dot\,)$, where
\beq
H_1=\2{\rm diag}(1,-1,-1,1)
\quad\hbox{and}\quad
H_2=\2{\rm diag}(0,1,0,-1)
\label{9.14}
\eeq
These vectors dual to the primitive roots are
\beq
W_1=\2{\rm diag}(1,0,-1,0)
\quad\hbox{and}\quad
\oW_2=\2{\rm diag}(1,1,-1,-1).
\label{9.15}
\eeq
Any of the properties a.), b.), or c.) of Sec. 2 shows that only
$\oW_2$ is minimal: For example, only $\oW_2$ exponentiates into the non-trivial element $(-\II)$ of Sp$(4)$: 
$$\exp2\pi W_1=1,
\qquad
\exp2\pi\oW_2=-\II.
$$
 In other words, while $W_1$ is already a charge, $\oW_2$ is only half-of-a-charge. (Fig. \ref{Figure7}).
%%%%%%%%%%%%%%%%%%%%%%%%%%
%%%%%%%%%%%%%%%%%%%%
Alternatively, the two remaining positive roots are $\varphi=\alpha_1+\alpha_2$ and the highest root is $\theta=2\alpha_1+\alpha_2$. 

Let the integer $m$ label the topological sectors. For $m$ even,
$m=2k$, the unique stable monopole belongs to the center,
\beq
\oQ^{(2k)}=k\,\Psi
\label{9.16}
\eeq
where $\Psi$ is a generator of the center normalized so that
$2\Psi$ is a charge. For $m$ odd, $m=2k+1$, the unique stable monopole is rather
\beq
\oQ^{(2k+1)}=({k+\2})\Psi+\2\oW_2\,.
\label{9.17}
\eeq
It may be worth noting that, in contrast to the $K=\SU(N)$ case,
$Q=\2W_1$ is an unstable monopole in the vacuum sector, which has index $2(\theta(W_1)-1)=2$ \footnote{Remark that if $W_1$ was the
charge of a Prasad-Sommerfield monopole, it would be stable \cite{24}.}.

The negative modes are expressed once more by (\ref{9.4}), but this time
$\sigma_{\pm}$ mean rather
\beq
\sigma_+=\2\barr{cccc}
0&1&&\\
0&0&&\\
&&0&-1\\
&&0&0
\earr,
\qquad
\sigma_-=\2\barr{cccc}
0&0&&\\
1&0&&\\
&&0&0\\
&&-1&0
\earr\ .
\label{9.18}
\eeq
\goodbreak
\kikezd{Acknowledgment}
We are indebted to J.-M. Souriau for hospitality in
Marseille where part of this work has been completed. We would also like to thank
Professors S. Coleman, P. Forgacs, Z. Horvath, J. Jones, D. Olive, L. Palla, A. Pressley,
D. Simms, and A. Wipf for discussions and correspondence.

\kikezd{Note added}. 
After this work has been completed, we heard from Prof. K. Uhlenbeck
that she and W. Nahm have obtained similar results \cite{28}.

\kikezd{Appendix}.
\kikezd{Proposition}
$$
m^2(\ba,\ba)=\int drd\Omega\tr(r\nabla_r\ba)^2=\smallover1/4+
\delta^2
\quad\hbox{and}\quad
{\rm inf\,}\delta^2=0.
$$
\noindent{\it Proof:}
\begin{eqnarray*}
\int_0^\infty drr^2\tr(\p_r\ba)^2
&=&
\int_R^\infty dr\tr(r\nabla_r\ba+\2\ba)^2
-
\int_R^\infty dr\tr(r\ba\nabla_r\ba)-
\smallover1/4\int_R^\infty dr\tr({\ba}^2)
%%%%%
\\[12pt]
&=&\smallover1/4\int_R^\infty dr\tr(\ba^2)+
\int_R^\infty dr\tr(r\nabla_r\ba+\2\ba)^2
+\frac{R}{2}\tr(\ba^2(R)).
\end{eqnarray*}
Therefore, $m^2=\smallover1/4+\delta^2$, as stated.

Equality can never be achieved, because $r\nabla_r\ba+\ba/2=0\,\Rightarrow\,\ba$ is proportional to $r^{-1/2}\,\Rightarrow
R\ba^2(R)\neq0$. However, consider $\ba=f(r)\bbeta(\Omega)$,
where $\bbeta(\Omega)$ is a vector on $\IS^2$, and
$$
f(r)=\left\{\begin{array}{ll}
(r-R)/R &R\leq r\leq 2R
\\
2(R/r)^{1/2} & 2R\leq r\leq 2sR
\\
s^{1/2}R &2sR\leq r
\end{array}\right..
$$
Then 
$$
\displaystyle\frac{\displaystyle\int\tr(r\nabla_rf)^2dr}
{\displaystyle\int f^2dr}=\frac{17+3\ln s}{5+12\ln s}
\to\smallover1/4
$$
as $s\to\infty$, showing that $\smallover1/4$ is indeed the infimum.

\newpage
%%%%%%%%%%%%%%%%%%
%\section{References}
%%%%%%%%%%%%%%%%%%

%%%%%%%%%%%%%%%%%%%%%%%%%%%%%%

\newpage

\kikezd{Note added in 2009: \textit{Hommage to Lochlainn
O'Raifeartaigh}
and to  \textit{Sidney Coleman.}}

Our aim in posting this  paper to ArXiv has been
two-fold. Firstly, we wanted to make available to a larger
public what,
22-years after its publication,  we still consider as one of our best papers we ever wrote.

 But it is also
\emph{Hommage} to two outstanding physicists, who played an important role in our personal history.

 Firstly, to 
\emph{Lochlainn
O'Raifeartaigh}, with whom we both (PAH and JHR) collaborated,
and from whom we had learned physics during those years we spent in Dublin. 

Lochlainn used to arrive at the DIAS around ten in the morning; coming directly to the kitchen.
While having coffee, he
pulled from his pocket an enveloppe with
some calculation on its back: ``I made some progress in the bus". Then the discussion started and went on for hours.
In fact, we spent more time
working on the tiny blackboard of the kitchen
than in the discussion room;  we published 
joint papers with him without having ever entered his office!

He had  a tremendous flair, picking the right idea
as a needle from a haystack.

He has also been a perfect gentlemen, whose collaborators ranged from the age of 25 to 70
(or beyond). 
And DIAS has deserved to be called \emph{School of Physics}.

This post is also an \emph{hommage} to \emph{Sidney Coleman}, with whom we
had less personal contacts, but who has, nevertheless, deeply influenced our work.

Sidney Coleman has been a true magician: 
in his celebrated Erice Lectures \cite{2} he could convey understanding, 
in a few words, to everyone, which no-one else could explain in dozens of  pages.  
He also had a tremendous intuition and a sparkling sense of humour. 

An illustration: in his '$81$ Erice Lecture  he claims
that \emph{``every topological sector contains exactly one stable
monopole charge"} --- but he only proves his statement in the
most trivial particular case, namely that of 
residual symmetry group $\SO(3)$ --- which almost 
never arises in physical applications: the most common examples
are rather of the GUT type, $SU(3)_c\otimes SU(2)_W\otimes U(1)_{em}$, with a continuous center.

His statement puzzled and angered us, and we wanted to find a general proof, valid for any compact Lie group. And after about a year of hard work, we 
realized that the problem could in fact be solved by  \dots{} 
factoring out the center and decomposing the resulting
semisimple group into $\SO(3)$ factors!

We were so fascinated by this ``coincidence'' that we wrote to Sidney Coleman,
asking if he was aware that his over-simplified idea contained in fact the germ of the general proof! His answer has also been
typical: --- ``I do not remember any more what I was aware 
of by that time;  you had better phone David Olive
or Werner Nahm who can tell you what I knew by that time!''

Another proof of his amazing intuition: in 
 his Lectures, Coleman mentions that \emph{the decay of monopoles
is analogous to the way elastic strings shrink}. 
In our paper, we have
been able to make this idea rigorous, and work out
the analogy between monopole decay with the energy-minimizing
shrinking of loops in the residual group!

Interest in monopoles in general, and in their stability in particular, has by now faded;  see, however Refs. \cite{Weinberg2006, GuoWe}. 
 
Our  original idea has been  that
our \emph{energy-reducing two-spheres} might indicate
preferential
\emph{decay routes} for an unstable monopole: starting
from some given unstable configuration, it would ``roll
down'' following these ``routes" to some lower-energy, 
``less-unstable'' saddle point, producing a sort of ``cascade''
of decaying monopoles ending eventually in the vacuum.

22 years ago, it was not
possible to check this intuitive picture. 
However, powerful computers and advanced numerical methods, unavailable in the past, might make it possible to test it today.

Both of these shining stars of our younger years are now dead. But we
want to reiterate our gratefulness for having been able to learn from them.
We include therefore, as an {\it Hommage}, photos of both 
of these - so different! - people.
\begin{figure}
\begin{center}
\includegraphics[scale=.4]{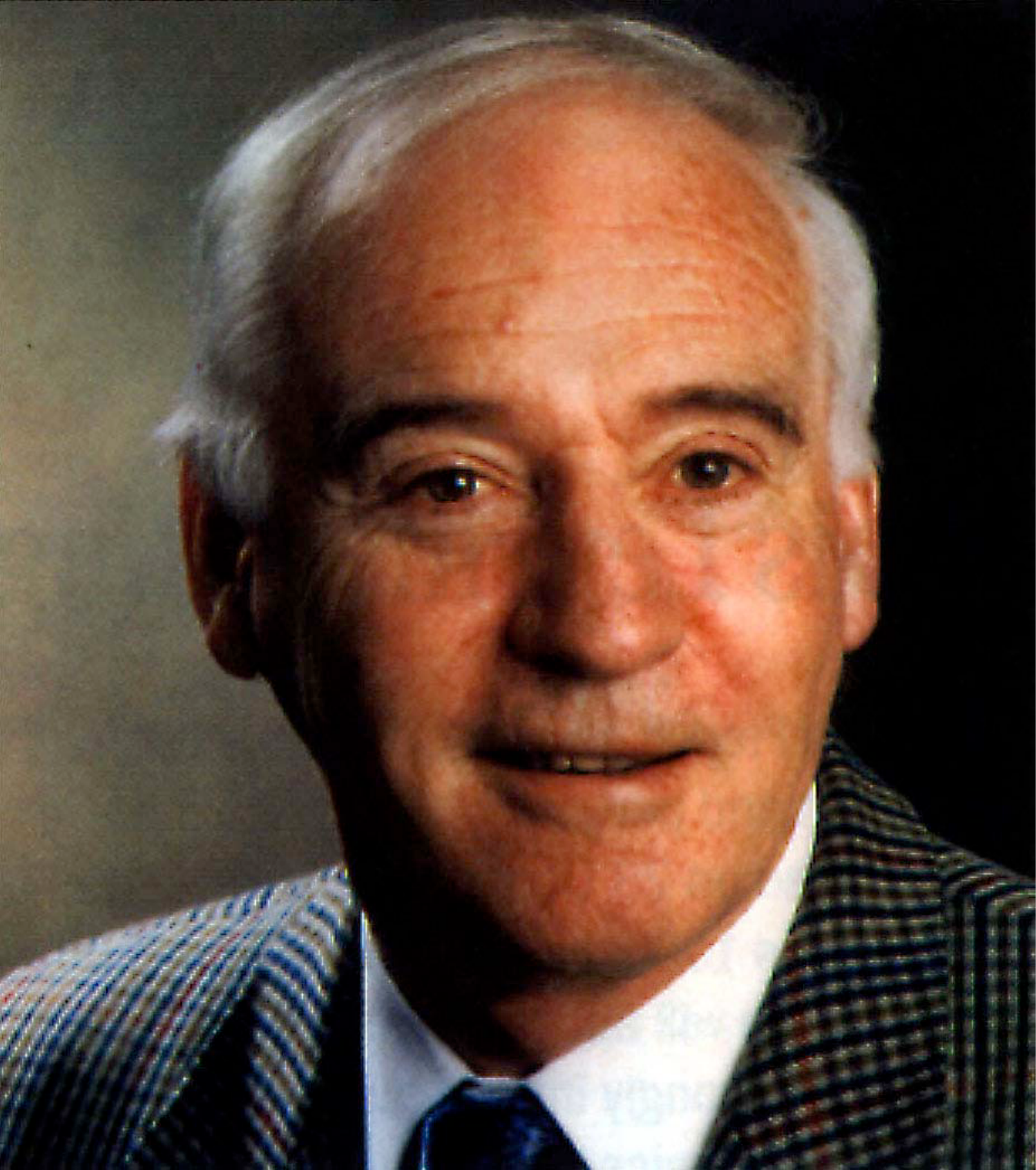}\qquad
\includegraphics[scale=1.1]{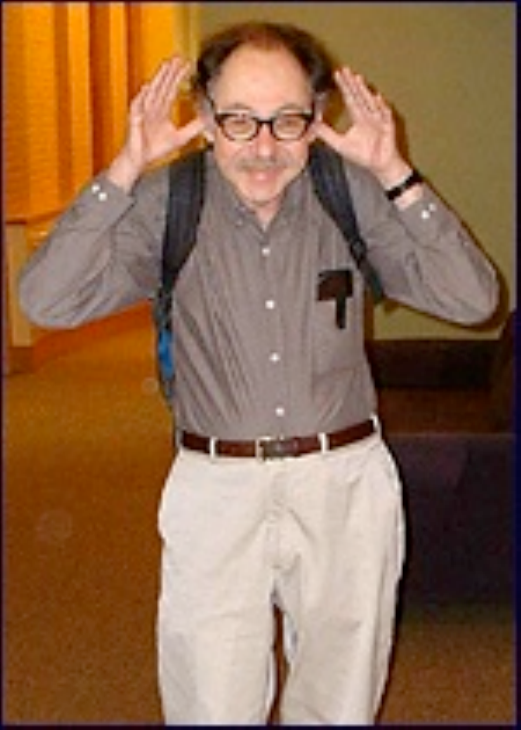}
\caption{Lochlainn O'Raifeartaigh and Sidney Coleman.}
\end{center}
\label{photos}
\end{figure} 

For further information on Sidney Coleman see:
http://www.physics.harvard.edu/QFT/.

Last but not least, we are indebted to Roman Jackiw, 
Stephen Parke, Rob Pisarski and Andreas Wipf for
correspondance, and for providing us with the
photos here.

\end{document}